\documentclass[12pt, a4paper,dvipdfmx]{article}
\usepackage{graphicx}
\usepackage{amssymb}
\usepackage{amsmath}
\usepackage{bm}
\usepackage{color}
\usepackage{cite}
\usepackage{epstopdf}            
\usepackage{epsfig}
\usepackage{axodraw4j}            

\newcommand{\beq}{\begin{equation}}
\newcommand{\eeq}{\end{equation}}

\newcommand{\gsim}{ \mathop{}_{\textstyle \sim}^{\textstyle >} }
\newcommand{\lsim}{ \mathop{}_{\textstyle \sim}^{\textstyle <} }
\newcommand{\sla}[1]{\not \! \! #1}

\usepackage[colorlinks=true, linkcolor=black, citecolor=black,
urlcolor=black]{hyperref}

\begin{document}

\begin{titlepage}

\begin{flushright}

\end{flushright}

\vskip 2cm
\begin{center}

{\Large
{\bf 
Effective theory approach \\to direct detection of dark matter
}
}

\vskip 2cm

Junji Hisano

\vskip 0.5cm

{\it 
Kobayashi-Maskawa Institute for the Origin of Particles and the
Universe, \\ Nagoya University, Nagoya 464-8602, Japan}\\[3pt]
{\it Department of Physics,
Nagoya University, Nagoya 464-8602, Japan}\\[3pt]
{\it 
Kavli IPMU (WPI), UTIAS, The University of Tokyo, Kashiwa, \\ Chiba 277-8583, Japan}

\vskip 1.5cm

\begin{abstract}

An effective field theory approach is presented for evaluation of the dark matter direct detection rate in this lecture note. This is prepared for the Les Houches Summer School Effective Field Theory in Particle Physics and Cosmology, July 2017.

\end{abstract}

\end{center}
\end{titlepage}

\section{Introduction}

There is now no doubt that dark matter (DM) exists in the
universe. However, we do not know the nature of the DM, since our
knowledge about DM is limited to the gravitational
aspect. We have no DM candidates in the standard model (SM) of particle
physics and also in astronomy, and the DM is now one of the
big issues in physics. The idea that DM may be
unknown particles produced in the early universe is fascinating, and many
candidates for the DM have been proposed \cite{Bertone:2004pz}.
Weakly-Interacting Massive Particles (WIMPs) are one of the leading
candidates. They are assumed to be produced in the thermal bath in the
early universe. The typical WIMP mass scale is $O(100)$~GeV to $O(1)$~TeV
under this assumption. We also expect new physics beyond the SM at the TeV scale from the naturalness point of view. This
coincidence is called the WIMP miracle. Many experiments now search
for WIMP DM. Direct detection of WIMP DM on Earth is
one of the methods. The WIMPs are assumed to pass through us. For
example, about one million WIMPs may exist in this room. Their interactions
are very weak, though there is a small probability that they may collide with nuclei. Direct detection experiments observe the recoiled
nuclei. Many such experiments are currently working or have been
proposed. A recent review of direct detection experiments is, for example,
given in Ref.~\cite{Undagoitia:2015gya}.

In this lecture, the WIMP DM detection rate from UV
theories at the TeV scale is evaluated. In this evaluation, the effective theory
approach works well. UV theories provide the interactions of WIMPs with
partons. On the other hand, we need to know the effective interactions of
WIMPs with nuclei. We have to derive effective theories at the parton, nucleon, and nuclei levels. In this lecture, we note the QCD
aspects in the evaluation of WIMP interactions with nucleons. We will show that we
can handle QCD corrections to the Wilson coefficients in the effective
interactions at the parton level well. We may then evaluate the
next-leading order contribution of $\alpha_s$, and the
strategy for the evaluation will be shown.

The lecture notes are organized as follows. First, we give
an introduction to the WIMP DM. After discussing the effective interactions of
WIMPs with nuclei and nucleons in Section~\ref{sec:2}, we give a brief review of the
direct detection experiments in Section~\ref{sec:3}. Then, we will show how
to evaluate the effective interactions of WIMPs with nucleons from UV
theories in Section~\ref{sec:4}. Due to the nucleon matrix elements of the
parton-level effective operators, the power counting of $\alpha_s$ in
calculating the direct detection rate is not the same as in
conventional ones. Furthermore, we do not necessarily need to
evaluate the Wilson coefficients for parton-level effective operators
at the hadronic scale with renormalization-group (RG) equations, in contrast
to the hadronic observables in flavor physics. These topics are discussed
in this section. In Section~\ref{sec:5} we show some results for three UV
models as examples: 1) gauge singlet WIMPs coupled with the Higgs boson, 2) gauge singlet
WIMPs coupled with colored scalars and quarks, and 3)
SU(2)$_L$ non-singlet WIMPs. We assume that the WIMPs are Majorana
fermions in this lecture. The application to other WIMPs is straightforward.
Finally, we discuss the strategy to evaluate the
direct detection rate including the $O(\alpha_s)$ correction in
Section~\ref{sec:6}. Section~\ref{sec:7} is devoted to the Summary. In
Appendix~\ref{FSgauge} we introduce Fock--Schwinger gauge fixing,
which is quite useful in evaluating the Wilson coefficients of the effective
operators including gluon field strengths. This lecture is mainly
based on the author's recent works \cite{Hisano:2015bma,Hisano:2015rsa}.

\section{WIMP DM}
\label{sec:1}

If the DM is composed of unknown particles, they are electrically
neutral. They are stable, or have a longer lifetime than the age of
the universe. They are massive so that they are ``cold'' in the structure
formation era of the universe, i.e., the free streaming length
after production in the early universe is shorter than the size of
protogalaxies so that the small-scale structure in the universe is not
erased. The cold DM abundance is precisely determined from CMB
power spectrum measurements. The DM particles are nonrelativistic in
the current universe, and the energy density $\rho_X$ is given by
$M_Xn _X$, with $M_X$ and $n_X$ the DM particle mass and number
density, respectively. It has been found from the CMB measurements that the
DM energy density normalized by the critical density in the universe,
$\Omega_X\equiv \rho_X/\rho_{\rm critical}$, is about $27\%$
\cite{Ade:2013zuv}. The critical density is $\rho_{\rm
 critical}\simeq 10^{-5}{\rm GeV/cm^2}$. The abundance and the free
streaming length depend on the production mechanism in the early
universe. In WIMP scenarios, the WIMPs are assumed be in
thermal equilibrium in the early hot universe.

One of the representative models for WIMPs is the supersymmetric
standard model (SUSY SM) \cite{SUSYDM}. In this model, a $Z_2$
symmetry called the $R$-parity is introduced, in order to stabilize
protons. As the result, the lightest SUSY particle is stable.
The neutral components of the fermionic superpartners of the gauge and Higgs
bosons, called gauginos and Higgsinos, are WIMP candidates in
the SUSY SM. Another representative model is the universal extra
dimension (UED) model \cite{Appelquist:2000nn}. In this model, we can impose
a parity symmetry in extra dimensional space, and the lightest
Kaluza--Klein particle is stable \cite{Cheng:2002ej}. The candidate in
the minimal model is the Kaluza--Klein photon. The SUSY SM and the UED model
are motivated by the naturalness problem in the Higgs boson mass
term in the SM so that their energy scale is expected to be at the TeV
scale. As will be explained below, the WIMPs have masses of about
$O(100)$~GeV-$O(1)$~TeV if they were produced in the thermal bath in the early
universe. These two observations support the assertion that new physics will appear
at the TeV scale. Many models have been proposed in order to explain
the naturalness and the WIMP DM.

Now, we evaluate the WIMP DM abundance in the universe. In the WIMP scenarios, the
WIMPs have interactions with the SM particles so that the WIMPs are thermalized
in the early hot universe. The stability of the
WIMPs comes from global symmetries, as given in the above examples. In
the early universe, where the temperature ($T$) is much higher than
the WIMP mass ($M_X$), they are in thermal equilibrium, and the number
density is comparable to those for the SM particles. When the temperature
decreases to below the WIMP mass, WIMP pair production by
SM particle collisions is suppressed in the thermal bath, so that the WIMP number
density deviates from that in thermal equilibrium. The WIMP
pair annihilation is frozen when the WIMPs do not find partners for
their pair annihilation within a Hubble time, and the number density is
only diluted by the expansion of the universe. Thus, the current
abundance of WIMPs is determined by the WIMP pair annihilation cross
section.

The WIMP abundance is more precisely evaluated with the Boltzmann
equation \cite{boltmaneq},
\begin{eqnarray}
 \frac{d n_X}{dt}+3H(T) n_X&=&-\langle \sigma|v| \rangle \left[n_X^2-(n_X^{\rm EQ})^2\right].
\end{eqnarray}
The second term is for dilution due to the expansion of the
universe. The Hubble parameter $H(T)$ in the radiation-dominated (RD) era
is given by
\begin{eqnarray}
 H(T)&=&\sqrt{\frac{8\pi}{3M_{\rm pl}}\rho}\simeq \sqrt{\frac{4\pi}{45}}g_\star^{1/2}
 \frac{T^2}{M_{\rm pl}}
\end{eqnarray}
where $M_{\rm pl}$ is the Planck mass and $\rho$ is the energy density
($\rho= (\pi^2/30)g_\star T^4$ with $g_\star=\sum_{\rm
 boson}1+\sum_{\rm fermion}7/8$). If the collision term in the
right-hand side of the Boltzmann equation is zero, $n_X/s$
($s$ is the entropy density, $s=(2\pi^2/45)g_\star T^3$) is constant
since $s$ is also diluted by the expansion of the universe.
In the collision term, $\langle \sigma|v|
\rangle$ is the thermal-averaged WIMP pair annihilation cross section
and $n_X^{\rm EQ}$ is the WIMP number density in thermal equilibrium.
The collision term is proportional to the square of the WIMP number
density since it comes from the WIMP pair annihilation and production.

The Boltzmann equation is rewritten as
\begin{eqnarray}
 \frac{x}{Y_{\rm EQ}}\frac{d Y}{d x}&=&-\frac{\Gamma_A}{H}\left[
   \left(\frac{Y}{Y_{\rm EQ}}\right)^2-1\right]
\end{eqnarray}
by defining $Y$ and $x$ as $Y\equiv n_X/s$ and $x=M_X/T$, respectively. Here, $\Gamma_A$
is the probability of annihilation per unit time for a WIMP,
\begin{eqnarray}
 \Gamma_A&=&n_X^{\rm EQ} \langle \sigma|v| \rangle.
\end{eqnarray}
When $x\gsim 1$, the WIMP pair production is kinematically suppressed,
and they start to be decoupled from the thermal bath so that $Y/Y_{\rm
 EQ}\gsim 1$. When $H\gg\Gamma_A$, the annihilation is frozen and $Y$
becomes constant. The freeze-out temperature ($T_F$) and the WIMP number
density at $T_F$ ($n_X^F$) are approximately determined by
$H(T_F)=\Gamma_X$, and it is found that $T_F$ is about $M_X/20$ and
$n_X^F\simeq H(T_F)/\langle \sigma|v| \rangle$. Thus, $Y$ and
$\Omega_X$ are approximately evaluated as
\begin{eqnarray}
 Y&\simeq&\sqrt{\frac{45}{\pi}}g_\star^{-1/2}\frac1{T_F M_{\rm pl}}\frac1{\langle \sigma|v| \rangle}, \nonumber \\ 
 \Omega_X&=&\frac{s^{\rm now}}{\rho_{\rm critical}} M_X Y
 \simeq 0.4\times
 \left(\frac{x_F\equiv M_X/T_F}{20}\right)
 \left(\frac{\langle \sigma|v| \rangle}{10^{-9} {\rm GeV}^{-2}}\right)^{-1},
\end{eqnarray}
where $s^{\rm now}$ is the entropy density in the current universe ($s^{\rm now}\simeq 3000{\rm cm^{-3}}$).

Let us discuss some typical cases. When the WIMPs are SU(2)$_L$ singlet
fermions, the annihilation cross section into SM fermions is given
by $\sigma v\sim \pi \alpha^2M_X^2/M_S^4$, with $M_S$ the mediator
scalar mass. Assuming that the mediator coupling constant $\alpha$ is
the same as that of the U(1)$_Y$ gauge interaction, $\sigma v\sim
3\times 10^{-9}{\rm GeV}^{-2}$ for $M_X=M_S=300{\rm GeV}$. If the
mediator mass is heavier than the WIMP mass, the cross section is
suppressed by $(M_X/M_S)^4$. If the WIMPs are Majorana fermions, the
annihilation into SM fermions suffers from $p$-wave suppression so
that the thermally averaged cross section is more suppressed by
$T_F/M_X\sim 1/20$ or the square of the masses of the SM fermions in the final
states. Thus, if the WIMPs are SU(2)$_L$ singlet Majorama fermions,
the WIMP mass is around 100~GeV. Binos (the fermionic superpartners of the
U(1)$_Y$ gauge boson in the SUSY SM) are an example. They are
SU(2)$_L$ singlet Majorana fermions. This situation may be changed
when some new particles are degenerate with the WIMPs in mass so that co-annihilation occurs \cite{Edsjo:1997bg}. For
example, if staus, which are the bosonic superpartners of the tau lepton, are
degenerate with the binos in mass, heavier binos are predicted due
to their co-annihilation.

On the other hand, if the WIMPs are the neutral component of the SU(2)$_L$
multiplet(s), they annihilate into two weak gauge bosons. The
annihilation cross section is approximately given by $ \sigma v\sim
\pi\alpha_2^2/M_X^2=3.5\times 10^{-9}~{\rm GeV}^{-2}\times (M_X/1~{\rm
 TeV})^{-2}$. Thus, the WIMP mass is expected to be at the TeV scale.
Higgsinos, which are the fermionic superpartners of Higgs bosons in the
SUSY SM, are SU(2)$_L$ doublets, and winos, which are those of
SU(2)$_L$ gauge bosons, are SU(2)$_L$ triplets. Detailed calculations
show that the SU(2)$_L$ doublet and triplet fermion masses are about
1~TeV and 3~TeV, respectively, if they are in thermal equilibrium in
the early hot universe \cite{Hisano:2006nn}.

\begin{figure}
\begin{center}
\includegraphics[width=6cm]{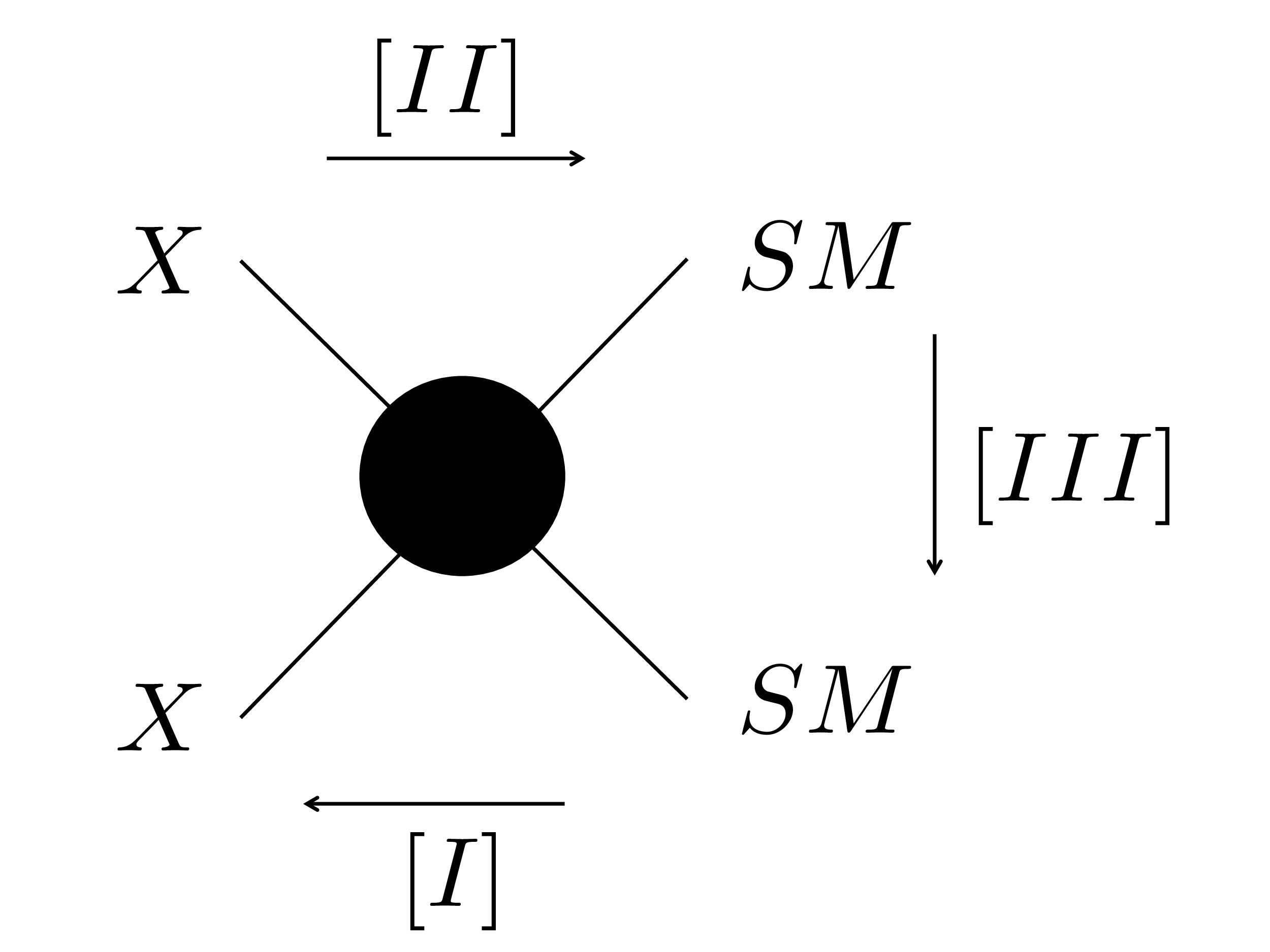}
\end{center}
\caption{Three approaches to WIMP searches.
\label{fig1}}
\end{figure}

Many kinds of experiments are currently searching for WIMPs in addition to signature
of new physics at the TeV scale. The first are direct searches for
WIMPs in collider experiments, such as in the LHC. (See $[I]$ in
Fig.~\ref{fig1}.) While WIMPs do not leave any signatures in detectors
in the experiments, their momenta appear missing in their events.
LHC experiments are now searching for events with missing transverse
momenta. The production cross sections of colored particles are
larger than the those for WIMPs at the LHC, and WIMPs are mainly
produced from the decay of the colored particles if they have interactions.

The second approach is indirect detection of WIMP DM in cosmic rays
($[II]$ in Fig.~\ref{fig1}). WIMPs are gravitationally
accumulated in massive astrophysical objects, such as stars, galaxies,
and galactic clusters. The WIMP pair annihilation is increased around
these objects since the annihilation rate is proportional to the square of
the WIMP density. The final states in the annihilation include gamma
rays, positrons, anti-protons, neutrinos, and so on, and they
contribute to cosmic rays. The Fermi satellite is observing gamma
rays, and it gives constraints on WIMP annihilation from observations of gamma rays
from the galactic center \cite{TheFermi-LAT:2015kwa} or dwarf galaxies
\cite{Ahnen:2016qkx}. The atmospheric Cherenkov telescopes, such as HESS and
MAGIC, are also searching for gamma rays from WIMP
annihilation \cite{Abramowski:2011hc}. AMS-02 on the International Space Station is searching for
antiparticles \cite{Aguilar:2016kjl}. The ICECUBE \cite{Aartsen:2012kia} and Super-Kamiokande experiments \cite{Choi:2015ara} observe solar
neutrinos to place constraints on WIMP annihilation in the
Sun. If non-standard signatures are found in cosmic rays, they may be
interpreted as WIMP annihilation. The indirect detection is
sometimes challenging due to the astrophysical backgrounds, though it
may be sensitive to heavy WIMPs with masses above the TeV scale. In particular,
the annihilation cross sections of SU(2)$_L$ non-singlet WIMPs are enhanced
by the attractive force in weak interactions, called the
Sommerfeld effect \cite{Hisano:2003ec}.

The third one is direct detection of WIMP DM trapped within our galaxy
($[III]$ in Fig.~\ref{fig1}). In these experiments, recoiled nuclei are
observed in elastic scattering with WIMPs, $X+{\rm N}\rightarrow
X+{\rm N}$. The typical recoil energy is $\sim m_r^2/m_T v^2$ ($m_r(=
m_T M_X/(m_T +M_X)$: reduced mass, $m_T$: target nucleus mass,
and $v$: DM velocity in the lab flame). The typical DM velocity
around the Earth is about $10^{-3}c$ ($c$: speed of light). Thus, the
recoil energy is $(1-100)$~keV. Currently, 1 ton-class detectors, such as
XENON1T \cite{Aprile:2017iyp} and PandaX-II \cite{Cui:2017nnn}, are
searching for WIMP DM. We will review the
experiments after showing the effective interaction of the nucleus/nucleon
with WIMPs.

\section{Effective WIMP-Nucleus/Nucleon Interactions}
\label{sec:2}

Let us discuss the WIMP-nucleus effective interactions in
the non-relativistic limit for  elastic scattering processes,
\begin{eqnarray}
X(\vec{p}) +{\rm N}(\vec{k})&\rightarrow& X(\vec{p'}) +{\rm N}(\vec{k'}),
\end{eqnarray}
where $\vec{p}$ and $\vec{p'}$ ($\vec{k}$ and $\vec{k'}$) are the incoming
and outgoing WIMP (nucleus, ${\rm N}$) three-momenta,
respectively. We assume that the WIMPs have spin $S_X$.
This discussion also applies to scalar and
vector WIMPs. The effective interactions are given by operators that
are constructed with four vectors, the momentum transfer
$\vec{q}(\equiv \vec{k}-\vec{k'})$, the incoming WIMP velocity
$\vec{v}$, the nucleus spin $\vec{S}_{\rm N}$, and the WIMP spin
$\vec{S}_X$. For convenience, we introduce $\vec{P}(\equiv\vec{p}+\vec{p'}= 2
\vec{p}+\vec{q})$ as an alternative to $\vec{v}$. In elastic
scattering, $|\vec{P}|$ and $|\vec{q}|$ are proportional to $v$
($v\equiv |\vec{v}|$).

The effective operators are classified into two categories: 
(nucleus) Spin-Independent (SI) and Spin-Dependent (SD) operators. 
There are four SI operators\cite{Fan:2010gt}:
\begin{eqnarray}
\begin{tabular}{ll}
$O_1^{(++)}=1$,& $O_2^{(-+)}=\vec{S}_X\cdot i \vec{q}$,\\
$O_3^{(--)}=\vec{S}_X\cdot\vec{P}$,~~~~~& $O_4^{(++)}=\vec{S}_X\cdot (\vec{P}\times i \vec{q})$.
\end{tabular}
\end{eqnarray}
Here, the superscripts ($\pm\pm$) are for the transformation properties
under parity ($P$) and charge conjugation ($C$). The SD
operators are the following: \begin{eqnarray}
\begin{tabular}{ll}
$O_5^{(++)}=\vec{S}_X\cdot\vec{S}_{\rm N}$,& $O_6^{(-+)}=\vec{S}_{\rm N}\cdot i \vec{q}$,\\
$O_7^{(--)}=\vec{S}_{\rm N}\cdot\vec{P}$, & $O_8^{(--)}=(\vec{S}_X\times\vec{S}_{\rm N})\cdot i \vec{q}$,\\
$O_9^{(-+)}=(\vec{S}_X\times\vec{S}_{\rm N})\cdot \vec{P}$,& $O_{10}^{(++)}
=\vec{S}_{\rm N}\cdot (\vec{P}\times i \vec{q})$,\\
$O_{11}^{(++)}=(\vec{S}_{\rm N}\cdot\vec{q})(\vec{S}_X\cdot \vec{q})$,~~~~~
& $O_{12}^{(++)}
=(\vec{S}_{\rm N}\cdot \vec{P})(\vec{S}_X\cdot\vec{P})$,\\
\multicolumn{2}{l}
{$O_{13}^{(+-)}=(\vec{S}_{\rm N}\cdot i\vec{q})(\vec{S}_X\cdot \vec{P})
+(\vec{S}_{\rm N}\cdot \vec{P})(\vec{S}_X\cdot i \vec{q})\nonumber$}\\
\multicolumn{2}{l}
{$O_{14}^{(+-)}=(\vec{S}_{\rm N}\cdot i\vec{q})(\vec{S}_X\cdot \vec{P})
-(\vec{S}_{\rm N}\cdot \vec{P})(\vec{S}_X\cdot i \vec{q})$}\\
\multicolumn{2}{l}
{$O_{15}^{(-+)}=
[\vec{S}_{\rm N}\cdot (\vec{P}\times \vec{q})](\vec{S}_X\cdot\vec{q})
+[\vec{S}_X\cdot (\vec{P}\times \vec{q})](\vec{S}_{\rm N}\cdot\vec{q})$}\\
\multicolumn{2}{l}
{$O_{16}^{(--)}=
[\vec{S}_{\rm N}\cdot (\vec{P}\times i \vec{q})](\vec{S}_X\cdot\vec{P})
+[\vec{S}_X\cdot (\vec{P}\times i \vec{q})](\vec{S}_{\rm N}\cdot\vec{P})$}.
\end{tabular}
\end{eqnarray}
The Wilson coefficients for these sixteen operators depend on
$|\vec{q}|$. The effective interactions depending on
$\vec{P}$ and/or $\vec{q}$ are suppressed by $v$. Two operators,
$O_1^{(++)}$ (SI) and $O_5^{(++)}$ (SD), are dominant in the elastic
scattering unless they are accidentally suppressed. In the following, 
we will concentrate on these cases. 

The effective interactions of WIMPs with nuclei are derived from those
with nucleons. When the WIMPs are Majorana fermions ($X$), the effective
interactions with nucleons ($N=n,p$) are simple, given 
by\footnote
{
While $X$ is a Majorana fermion, we omit ``1/2'' in the coefficients of the 
interactions due to the convention given in Ref.~\cite{Drees:1993bu}.
}
\begin{eqnarray}
{L}_{\rm eff}&=&
\sum_{N=n,p}
\left(
f_N \bar{X}X \bar{N}N 
+a_N \bar{X}\gamma^\mu\gamma_5 X \bar{N} \gamma_\mu\gamma_5 N 
\right).
\label{leff-necleon}
\end{eqnarray}
The first term is the SI interaction in the non-relativistic limit,
while the second one is the SD one. Neglecting nucleus form
factors, the elastic-scattering cross sections of WIMPs with nuclei are
given by
\begin{eqnarray}
\sigma=\sigma_{\rm SI}+\sigma_{\rm SD}
\end{eqnarray}
with
\begin{eqnarray}
\sigma_{\rm SI}&=&\frac{4}{\pi}{m_r^2}|Z f_p+(A-Z) f_n|^2,\nonumber\\
\sigma_{\rm SD}&=&\frac{16}{\pi}{m_r^2}\frac{J+1}J 
|a_p \langle S_p\rangle 
+a_n \langle S_n\rangle |^2.
\end{eqnarray}
Here, $Z$ and $A$ are the atomic and mass numbers, respectively, and $J$
and $\langle S_N\rangle$ are the total spin of target nucleus and
the expectation value of the nucleon spin in the nucleus, respectively. 

While nuclei with non-zero spin are sensitive to the SD interactions, 
nuclei with larger atomic or mass numbers are more sensitive to
the SI interactions. The nucleon scalar operators in Eq.~\ref{leff-necleon}
become proton and neutron density operators in the non-relativistic limit.
Thus, the amplitude is proportional to $Z$ or $A-Z$. 

Earlier, we ignored the nucleus form factors. However, they are not
negligible in practice. Let us consider the SI cross
section. The momentum transfer $q=|\vec{q}|$ is given by $(2m_T
E_R)^{1/2}$, where the recoil energy $E_R$ is $E_{th}\le E_R \le
2(m_r^2/m_T) v^2$. The nuclear radius $r_{\rm N}$ is typically
$A^{1/3}$~fm, and  then, $r_{\rm N} q\simeq 7\times 10^{-3} A^{5/6}
(E_R({\rm keV}))^{1/2}$. While the SI cross sections are larger for
larger nuclei, the SI cross sections suffer from suppression for large
$E_R$ and/or large $A$ due to the form factors. Furthermore, the form
factors are important to predict the precise spectrum of the nucleus
recoil energy in direct detection. The nucleus form factors are
reviewed in Ref.~\cite{Lewin:1995rx}.

\section{Brief Review of Direct Detection Experiments}
\label{sec:3}

The DM energy density around the earth is determined by the stellar motions in our galaxy. However, due to our limited knowledge of the DM density distribution in our galaxy, the energy distribution is still evaluated with a large uncertainty as \cite{Read:2014qva}
\begin{eqnarray}
\rho_X&=&(0.2-0.6){\rm GeV/cm^3}.
\end{eqnarray}
The DM velocity distribution is more uncertain since we have no way to measure it directly. A Maxwellian distribution $f(\vec{v})$ is assumed in the galactic center coordinate \cite{Lewin:1995rx},
\begin{eqnarray}
f(\vec{v})\simeq \frac1{(\pi v_0^2)^{3/2}} {\rm e}^{-(\vec{v}+\vec{\oplus})^2/v_0^2},
\end{eqnarray}
where $v_0\simeq 230{\rm km/sec}$ and $\vec{v}_{\oplus}$ is the Earth's velocity.
This distribution is supported by the $N$-body simulations.
The Earth's velocity is 
\begin{eqnarray}
|\vec{v}_{\oplus}|\simeq 244+ 15 \sin(2\pi y)~({\rm km/sec}),
\end{eqnarray}
with $y$ the elapsed time from March 2nd. The first term comes from
the motion of the Solar System, and the second is from revolution of the
Earth around the Sun. If WIMP wind relative to the Earth is detected by
direction-sensitive detectors, it would be strong evidence for the
DM, although it would still not be conclusive. Annual modulation of the
event rate is also expected, and it is at most 3\%.

Thus, the signals in the direct detection experiments are only from recoiled
nuclei. The signals are observed by detecting phonons/heat, ionization,
or light generated by the recoiled nuclei in the experiments. They have to
maintain a low background in the experiments. The dominant backgrounds
are gamma rays and electrons from the beta and gamma decays. The
experiments are performed underground, and the coincidence of the
signals, such as ionization and light in the XENON \cite{Aprile:2017iyp}
and PandaX experiments \cite{Cui:2017nnn}, are observed.

The event rate per unit mass of the target $R$ is given as
\begin{eqnarray}
 dR&=& \frac{N_0}{A}\sigma v~dn_X,
\end{eqnarray}
($N_0$ is the Avogadro number). Assuming a zero-momentum transfer cross section $\sigma\simeq {\rm const}=\sigma_0$, the event rate is evaluated as
\begin{eqnarray}
 R_0&=&\frac{2}{\pi} \frac{N_0}{A} \frac{\rho_X}{M_X}\sigma_0 v_0\nonumber\\
 &\simeq&\frac{540}{AM_X}
 \left(\frac{\sigma_0}{10^{-36}{\rm cm^2}}\right)
  \left(\frac{\rho_0}{0.4{\rm GeV/cm^3}}\right)
  \left(\frac{v_0}{230{\rm km/sec}}\right)
  ~{\rm event/kg/day},
\end{eqnarray}
where $M_X$ is in GeV.

The event spectrum as a function of the recoil energy $E_R$ is
approximately given by \cite{Lewin:1995rx}
\begin{eqnarray}
 \frac{dR}{dE_R}\simeq \frac{R_0}{E_0}{\rm e}^{-E_R/E_0}
\end{eqnarray}
where $E_0=2 (m_r^2/M_T)v^2$. Thus, if the energy threshold in the experiment is lowered, the event rate is increased and the experiments are more sensitive to WIMPs with lighter masses.

As explained in the previous section, the SI cross section of WIMPs
with nuclei is enhanced when the target nucleus has large atomic or
mass numbers. Thus, many experiments use heavy nuclei as the
targets. If the SI interaction is isosinglet, the
zero-momentum transfer SI cross sections with nuclei are proportional to the SI
cross section with the proton $\sigma^p_{\rm SI}$ as
\begin{eqnarray}
 \sigma_0&=& A^2\frac{m_r^2(A)}{m_r^2(1)} \sigma^p_{\rm SI}
\end{eqnarray}
with ${m_r^2(A)}$ $({m_r^2(1)})$ the reduced mass of the target nucleus
(proton) and the WIMP. The sensitivity and exclusion curves of the
direct detection experiments are shown on a plane of $\sigma^p_{\rm SI}$
and $M_X$.

The current limits on the SI cross section of proton are derived from
LUX \cite{Akerib:2016vxi}, XENON1T \cite{Aprile:2017iyp}, and
PandaX-II \cite{Cui:2017nnn}. They exclude up to $\sigma^p_{\rm SI}\sim
10^{-46}{\rm cm}$ for $M_X \simeq 50~{\rm GeV}$. The limits for heavier
WIMPs are scaled by $1/M_X$ since the DM number density is
$\rho_X/M_X$. The second-generation experiments, XENONnT, LZ, and
PandaX-xT, whose fiducial volumes are $O(1)$~ton, will start in a few
years \cite{Undagoitia:2015gya}. These experiments will aim for
$10^{-48}~{\rm cm^2}$. 

Neutrino coherent scattering off nuclei is a serious background source in
direct detection experiments, called ``the neutrino floor''
\cite{Gutlein:2010tq}.　It is quite difficult to remove it except in direct detection experiments with directional sensitivity. The solar
neutrinos hide the signals for light DM ($M_X\lsim O(1)~{\rm GeV}$) if
$\sigma^p_{\rm SI}\lsim 10^{-44}{\rm cm^2}$. The atmospheric neutrinos
also hide the heavier DM signals if $\sigma^p_{\rm SI}\lsim
10^{-49}~(10^{-48}){\rm cm^2}$ for $M_X\simeq 100$ GeV (1~TeV). 
The third-generation experiments aim to reach the neutrino
floor \cite{Undagoitia:2015gya}.

\section{Evaluation of Effective Interaction of WIMPs with Nucleons}
\label{sec:4}
We now evaluate the effective interactions of WIMPs with nucleons from
UV theories. For this purpose, we first construct effective theories of WIMPs
and quarks/gluons for direct detection experiments. The Wilson
coefficients $C_i(\mu_{\rm UV})$ ($i=1,2,\cdots$) of the effective
operators $O_i(\mu_{\rm UV})$ at the UV scale $\mu_{\rm UV}$ are
derived by integrating out heavy particles in the UV theory. In the normal
procedure for hadronic observables in flavor physics, the Wilson
coefficients at the hadronic scale $\mu_H\sim 1$~GeV are derived from
those at the UV scale using the RG equations. This is the case where
our knowledge about the matrix elements of the effective operators is
limited to those at the hadronic scale. However, some matrix elements
relevant to DM direct detection are available at any $\mu$ so that
we do not need to evaluate the Wilson coefficients at the hadronic
scale ourselves, especially at the leading order of $\alpha_s$.

In this section, we assume that the WIMPs are Majorana fermions, though the
derivation in this section is applicable to the scalar and
vector WIMPs, as given in Ref.~\cite{Hisano:2015bma}.

\subsection{Effective Theories at the Parton Level}

When the WIMPs are Majorana fermions, the parton-level effective
interactions relevant to DM direct detection at the hadronic scale
are the following:
\begin{equation}
 { L}_{\rm eff} = 
\sum_{p=q,g}C^p_S { O}^p_S
+\sum_{i=1,2}\sum_{p=q,g}C^p_{T_i}{ O}^p_{T_i}
+\sum_{q} C^q_{AV}{ O}^q_{AV}
~,
\label{parton-level-op}
\end{equation}
with
\begin{eqnarray}
 {O}^q_S&\equiv&
 \bar{X}X~
 m_q\overline{q}q~,\nonumber \\ 
 {O}^g_S&\equiv& 
 \bar{X}X~
 \frac{\alpha_s}{\pi}
 G^A_{\mu\nu}G^{A\mu\nu}_{}~,\nonumber\\ 
 { O}^p_{T_1}&\equiv& \frac{1}{M_X}
\bar{X}i\partial^\mu \gamma^\nu X ~{ O}^p_{\mu\nu}~,\nonumber\\
 { O}^p_{T_2}&\equiv& \frac{1}{M_X^2}
 \bar{X} i\partial^\mu i \partial^\nu
 X ~{ O}^p_{\mu\nu}~,\nonumber\\
 { O}^q_{AV}&\equiv&
 \bar{X}\gamma_\mu^{}\gamma_5^{}X~ \overline{q}\gamma^\mu_{}\gamma_5^{}
 q
~,
\label{parton-level-op2}
\end{eqnarray}
up to the equations of motions and the integration by parts. Here,
$q$ and $G^A_{\mu\nu}$ denote the light quarks ($q=u,d,s$) and the field
strength tensor of the gluon field, respectively; $m_q$ are the masses of the
quarks; $\alpha_s\equiv g_s^2/(4\pi)$ is the strong coupling constant,
${ O}^q_{\mu\nu}$ and ${ O}^g_{\mu\nu}$ are called the spin-2 twist-2
operators of the quarks and the gluon, respectively, 
\begin{eqnarray}\label{twist2op}
 { O}^q_{\mu\nu}&\equiv& \frac{1}{2}\overline{q}i\biggl(
D_\mu^{}\gamma_\nu^{} +D_\nu^{}\gamma_\mu^{}-\frac{1}{2}g_{\mu\nu}^{}
\sla{D}\biggr)q~,\nonumber \\
{ O}^g_{\mu\nu}&\equiv& 
G^{A\rho}_{\mu} G^{A}_{\nu\rho}-\frac{1}{4}g_{\mu\nu}^{}
G^A_{\rho\sigma}G^{A\rho\sigma}~,
\end{eqnarray}
with $D_\mu$ as the covariant derivatives. The quark or gluon field strength bilinear operators are up to dimension 4 in
Eq.~\ref{parton-level-op}. The operators $O_{AV}^q$ contribute to the
SD interactions, while the other operators contribute to the SI ones.

The quark/gluon scalar operators in $O_S^q$ and $O_S^g$ are multiplied
by $m_q$ and $\alpha_s/\pi$, respectively. The reasons are the
following. The quark scalar operators are chiral symmetry-breaking so
that they are multiplied by $m_q$. The nucleon matrix elements of
$G^A_{\mu\nu} G^{A\mu\nu}$ are multiplied by $\pi/\alpha_s$, compared
with those of $m_q\bar{q}q$, as will be shown. This leads to a change
in the power counting of $\alpha_s$ in the perturbation when
evaluating the DM direct detection rate. Then, we multiply $O_S^g$ by
$\alpha_s/\pi$. In addition, $O_S^q$ and $O_S^g$ are RG invariant at
all orders and at the one-loop level, respectively, in QCD. These will
be explained in next subsection in more detail.

Twist is defined as the mass dimension minus the spin of the
operators. Higher spin or higher twist operators have more mass
dimensions than the spin-2 twist-2 operators so they are negligible in
DM direct detection \cite{Drees:1993bu}. The operators in $O_{T1}^p$
and $O_{T2}^p$ ($p=q,g$) have mass dimensions 8 and 9, respectively,
higher than the others in Eq.~\ref{parton-level-op2}. However, in the
non-relativistic limit for WIMPs, $O_{T1}^p\simeq X^\dagger X
O^p_{00}$ and $O_{T2}^p\simeq \bar{X}X O^p_{00}$, which behave as
dimension-7 operators, since the operators in $O_{T1}^p$ and
$O_{T2}^p$ are multiplied by $1/M_X$ and $1/M_X^2$,
respectively\footnote{
This is more transparent in heavy particle effective theories, where
the WIMPs are treated as non-relativistic fields \cite{Hill:2013hoa}.
}
\cite{Drees:1993bu}.
 Unless the mediator particles
generating those operators have masses much larger than $M_X$, the
Wilson coefficients for $O_{T1}^p$ and $O_{T2}^p$ are not
suppressed. Thus, they may contribute to the SI interactions,
comparable to $O^q_S$ and $O^g_S$.

\subsection{Matrix Elements and RG Equations for Scalar Operators}

First, let us discuss the nucleon matrix elements for the quark/gluon
scalar operators. The matrix elements for the quark scalar operators in
$O_S^q$ are given by
\begin{eqnarray}
\langle N|m_q \bar{q}q|N\rangle\equiv f_{T_q}^{(N)} m_N,
\end{eqnarray}
with $f_{T_q}^{(N)}$ the mass fraction parameters of the quark $q$. Presently,
the mass fraction parameters of quarks are evaluated with lattice QCD
simulations. The following were derived by the ETM collaboration with
$N_f=2+1+1$ \cite{Abdel-Rehim:2016won},
\begin{eqnarray}
f_{T_u}^{p}=0.0149(17)(^{21}_{16}),& 
f_{T_u}^{n}=0.0117(15)(^{18}_{12}),\nonumber\\
f_{T_d}^{p}=0.0234(23)(^{27}_{16}),&
f_{T_d}^{n}=0.0298(23)(^{30}_{16}), \nonumber\\
f_{T_s}^{N}=0.0440(88)(^{72}_{15}),& \nonumber\\
f_{T_c}^{N}=0.085(22)(^{11}_{7}), &
\label{massfraction}
\end{eqnarray}
($N=p,n$). The first and second parentheses are for statistical and systematical
uncertainties, respectively. Heavier quarks have larger mass fraction
parameters.  In the simulations, the mass fraction parameter of the
charm quark is also evaluated. We will return to this later.

The nucleon matrix element of the gluon scalar operator is evaluated
with the trace anomaly in QCD. The trace anomaly in QCD is \cite{Shifman:1978zn}
\begin{eqnarray}
\theta^\mu_\mu&=&
\frac{\beta(\alpha_s)}{4\alpha_s}
G^A_{\mu\nu} G^{A\mu\nu}
+
(1-\gamma_m(\alpha_s)) \sum_q m_q \bar{q}q.
\end{eqnarray}
The beta function of $\alpha_s$, $\beta(\alpha_s)$, and anomalous dimension of the quark mass, $\gamma_m(\alpha_s)$,
are given by
\begin{eqnarray}
\beta(\alpha_s) &\equiv&\mu\frac{\partial }{\partial \mu}\alpha_s
\simeq 2 b_1\frac{\alpha_s^2}{4\pi}+2 b_2\frac{\alpha_s^3}{(4\pi)^2},
\nonumber\\
\gamma_m(\alpha_s) m_q &\equiv&\mu\frac{\partial}{\partial \mu} m_q
\simeq -6 C_F \frac{\alpha_s}{4\pi},
\end{eqnarray}
where $b_1=-11N_c/3+2N_f/3$ and $b_2=-34 N_c^2/3+10N_c N_f/3+2C_F N_f$ ($N_c=3$ and $C_F=4/3$).
The nucleon mass is given by the nucleon matrix element of the trace anomaly,
\begin{eqnarray}
m_N&\equiv&\langle N|\theta^\mu_\mu|N\rangle.
\end{eqnarray}
Thus, the nucleon matrix element of the gluon scalar operator is given by 
\begin{eqnarray}
\langle N|
\frac{\alpha_s}{\pi} G^A_{\mu\nu} G^{A\mu\nu}
|N\rangle
&=&
m_N\frac{4\alpha_s^2}{\pi \beta(\alpha_s)}\left[
1-(1-\gamma_m(\alpha_s))\sum_q f_{T_q}^{(N)}\right]\nonumber\\
&\simeq&
-\frac89 m_N(1-\sum_qf_{T_q}^{(N)})+O(\alpha_s). 
\end{eqnarray}
We take $N_f=3$ in the last of the above equations. When
the gluon scalar operator is multiplied by $\alpha_s/\pi$, the nucleon
matrix element is O(1) in units of $m_N$. 

In most UV models, the WIMPs do not directly couple with gluons,
and the effective interactions of the WIMPs with the gluon are generated by the
integration of the quarks or other colored particles. This implies that the
effective interaction for the gluon scalar operator is suppressed by
$\alpha_s/\pi$ compared to those for the quark scalar operators.
However, the nucleon matrix elements for the gluon scalar operators
are $O(1)$ even if the gluon scalar
operator is multiplied by $\alpha_s/\pi$. Thus, we have to evaluate the higher-order contributions to the
gluon scalar operator by $\alpha_s/\pi$, in contrast to the quark
scalar operators.

Let us discuss the contribution of the heavy quark scalar operators
$m_Q\bar{Q}Q \bar{X}X $ to the gluon scalar operator in order to see
the above counting of $\alpha_s$. The nucleon matrix element of the
trace anomaly is independent of the number of flavors $N_f$ since it is a
physical observable. This implies that
\begin{eqnarray}
\langle N| m_Q \bar{Q}Q|N \rangle
&=&
\frac{\pi \Delta\beta(\alpha_s)}{4(1-\gamma_m(\alpha_s))\alpha_s^2}
\langle N|
\frac{\alpha_s}{\pi} G^A_{\mu\nu} G^{A\mu\nu}
|N\rangle
\nonumber\\
&\simeq&
-\frac{1}{12}(1+11\frac{\alpha_s}{4\pi})
\langle N|
\frac{\alpha_s}{\pi} G^A_{\mu\nu} G^{A\mu\nu}|N\rangle
\nonumber\\
&\simeq&
\frac{2}{27}m_N,
\label{heavyquarkintegral}
\end{eqnarray}
at $\mu\simeq m_Q$. Here, $\Delta
\beta(\alpha_s)=\beta(\alpha_s)|_{N_f}-
\beta(\alpha_s)|_{N_f-1}$. Thus, if the Wilson coefficients for the heavy and
light quark scalar operators are common, the heavy quark contribution
via the gluon scalar operator dominates over the light quark ones.

It is found that the numerical value in Eq.~\ref{heavyquarkintegral}
is consistent with the mass fraction of the charm quark in
Eq.~\ref{massfraction}. This coincidence is welcome, though a more
precise determination of the mass fraction of the charm quark may reduce
the uncertainty in the predicted direct detection rate. The charm
quark mass is close to the hadronic scale so that the higher-dimensional operators might not be negligible after integrating out
the charm quark. By integrating out the heavy quarks $Q$, the dimension-6
operators are generated as \cite{Cho:1994yu}
\begin{eqnarray}
-\frac{\alpha_s}{12\pi} G^A_{\mu\nu} G^{A\mu\nu}
+\frac{\alpha_s}{64\pi m_Q^2}
(D^\nu G_{\nu\mu})^A (D_\rho G^{\rho\mu})^A
-\frac{g_s \alpha_s}{720m_Q^2}f_{ABC} G^{~~A}_{\mu\nu}G^{\mu\rho B} G^{\nu~C}_{~\rho}.
\nonumber\\
\end{eqnarray}
The first term corresponds to the leading order term in the second
line of Eq.~\ref{heavyquarkintegral}. When $\Lambda_{\rm
 QCD}^2/m_c^2\simeq O(10)\%$, the coefficients of the
higher-dimensional operators are numerically suppressed so that they are
expected to be a few \% of the leading term. This should be
justified in lattice QCD in order to derive reliable predictions about
the DM direct detection.

The anomalous dimensions of the scalar operators are also derived 
from the RG invariance of the quark scalar operators and the trace anomaly,
\begin{eqnarray}
\mu\frac{\partial}{\partial \mu} m_q\bar{q}q=0,&~&
\mu\frac{\partial}{\partial \mu} \theta^\mu_\mu=0.
\end{eqnarray}
It is found that
\begin{eqnarray}
\mu \frac{\partial}{\partial \mu}
\left(C_S^q,C_S^g \right)
&=&
\left(C_S^q,C_S^g \right)
\Gamma_S,
\end{eqnarray}
where $\Gamma_S$ is an $(N_f+1)\times(N_f+1)$ matrix,
\begin{eqnarray}
\Gamma_S
&=&
\left(
\begin{tabular}{cccc}
$0$ & $\cdots$ & 0&0\\
$\vdots$ & $\ddots$ & $\vdots$ & $\vdots$ \\
$0$ & $\cdots$ & $0$ & $0$ \\
$-\frac{4\alpha_s^2}{\pi}
\frac{\partial\gamma_m(\alpha_s))}{\partial \alpha_s}$
&$\cdots$&
$-\frac{4\alpha_s^2}{\pi}
\frac{\partial\gamma_m(\alpha_s)}{\partial \alpha_s}$
&$\alpha_s^2
\frac{\partial}{\partial {\alpha_s}}\left(\frac{\beta(\alpha_s)}{\alpha_s^2}\right)$
\end{tabular}
\right).
\end{eqnarray}
$C_S^q$ and $C_S^g$ are RG-invariant in the leading term
of $O(\alpha_s)$. The solutions for the above equations are
\begin{eqnarray}
C_S^q(\mu)&=&C_S^q(\mu_0)
-\frac4\pi C_S^G(\mu_0)(\gamma_m(\alpha_s(\mu))-\gamma_m(\alpha_s(\mu_0))),\nonumber\\
C_S^g(\mu)&=& \frac{\beta(\alpha_s(\mu))}{\alpha_s^2(\mu)}
\frac{\alpha_s^2(\mu_0)}{\beta(\alpha_s(\mu_0))}C_S^g(\mu_0).
\end{eqnarray}

\subsection{Matrix Elements and RG Equations for Twist-2 Operators}

Next are the matrix elements for the spin-2 twist-2 operators.
The matrix elements are given by the parton-distribution functions (PDFs)
of the nucleons $(N=p,n$) as
\begin{eqnarray}
\langle N(p)| {O}^q_{\mu\nu}(\mu)|N(p) \rangle
&=&
\frac1{m_N}\left(p_\mu p_\nu-\frac14 m_N^2 g_{\mu\nu}\right)
(q^{(N)}(2;\mu)+\bar{q}^{(N)}(2;\mu)),\nonumber\\
\langle N(p)| {O}^g_{\mu\nu}(\mu)|N(p) \rangle
&=&
-\frac1{m_N}\left(p_\mu p_\nu-\frac14 m_N^2 g_{\mu\nu}\right)
g^{(N)}(2;\mu),
\label{twist2matrixemement}
\end{eqnarray}
where $q^{(N)}(2;\mu)$, $\bar{q}^{(N)}(2;\mu)$, and $g^{(N)}(2;\mu)$
are the second moments of the PDFs for the quark, antiquark, and gluon, respectively.
The $n$-th moments of the PDFs are defined as 
\begin{eqnarray}
q^{(N)}(n;\mu)&\equiv&\int^1_0 dx~x^{n-1} q^{(N)}(x;\mu),\nonumber\\
\bar{q}^{(N)}(n;\mu)&\equiv&\int^1_0 dx~x^{n-1} \bar{q}^{(N)}(x;\mu),\nonumber\\
g^{(N)}(n;\mu)&\equiv&\int^1_0 dx~x^{n-1} g^{(N)}(x;\mu),
\end{eqnarray}
where $q^{(N)}(x;\mu)$, $\bar{q}^{(N)}(x;\mu)$, and $g^{(N)}(x;\mu)$
are the PDFs of the quark, antiquark, and gluon, respectively, at 
the factorization scale $\mu$. 

The derivation of Eq.~\ref{twist2matrixemement} is given in standard
textbooks of quantum field theory, such as texts by Peskin and Schroeder
\cite{Peskin:1995ev} or by Schwartz \cite{Schwartz:2013pla}. In the standard derivation,
Operator Product Expansions (OPEs) are applied to Deeply-Inelastic
Scattering (DIS), $e^-+N\rightarrow e^-+X$. In the expansion, the
twist-2 operators are dominant, and the higher-twist operators are
suppressed by the momentum transfer. From the contour integral of
the OPEs on the complex plane of $\omega=1/x$ ($x$: the Bjorken $x$ in PDFs),
it can be shown that
\begin{eqnarray}
\langle N(p)| {O}^q_{\mu_1\cdots\mu_n}(\mu)|N(p) \rangle
&=&
\frac1{m_N}\left\{p^{\mu_1}\cdots p^{\mu_n}\right\}_{\rm TS}
(q^{(N)}(n;\mu)+(-1)^n \bar{q}^{(N)}(n;\mu)),
\nonumber\\
&&
\label{pdf-twist-2}
\end{eqnarray}
where ${O}^q_{\mu_1\cdots\mu_n}(\mu)$ are twist-2 spin-$n$ operators
and ``TS'' means traceless symmetric. This derivation is for only 
the leading order term of Eq.~\ref{twist2matrixemement}. The $\mu$ dependence
is introduced through the radiative correction. It can be shown that
the anomalous dimensions for twist-2 operators are consistent with
the Altarelli-Parisi evaluation of PDFs. 

There is another derivation of Eq.~\ref{twist2matrixemement}. In the
above derivation using DIS, it is assumed that the integral along
$|\omega| \rightarrow \infty$ on the complex plane vanishes. If we
define the PDFs using quantum fields, we may derive
Eq.~\ref{twist2matrixemement} in a rigorous way without such an
assumption of a specific process. It has been proposed by Collins and Soper
\cite{Collins:1981uw} that the PDFs be defined in light-cone
coordinates. Eq.~\ref{twist2matrixemement} is derived directly from
the definition. Furthermore, Eq.~\ref{pdf-twist-2} is the Mellin
transformation of the PDFs in the mathematical language. Thus, the inverse
Mellin transformation of the matrix elements of twist-2 operators
gives another definition of the PDFs. It gives a basis to evaluate of the
PDFs in lattice QCD simulations \cite{Soper:1996sn}.

Several groups evaluated the PDFs of partons by fitting with measurements in
collider experiments, and they provide the PDFs at any factorization scale.
The following are the second moments of the PDFs of proton derived by the CTEQ-Jefferson Lab. collaboration
\cite{Owens:2012bv}, 
\begin{eqnarray}
g^{(p)}(2,\mu)=0.464(2), &&\nonumber\\
u^{(p)}(2,\mu)=0.223(3),&& \bar{u}^{(p)}(2,\mu)=0.036(2),\nonumber\\
d^{(p)}(2,\mu)=0.118(3),&& \bar{d}^{(p)}(2,\mu)=0.037(3),\nonumber\\
s^{(p)}(2,\mu)=0.0258(4),&& \bar{s}^{(p)}(2,\mu)=s(2,\mu),\nonumber\\
c^{(p)}(2,\mu)=0.0187(2),&& \bar{c}^{(p)}(2,\mu)=c(2,\mu),\nonumber\\
b^{(p)}(2,\mu)=0.0117(1),&& \bar{b}^{(p)}(2,\mu)=b(2,\mu),
\end{eqnarray}
where $\mu=m_Z$ and $N_f=5$. The second 
moments of valence quarks and the gluon are $O(1)$ and those of the
sea quarks are sub-leading, as expected. Those of the neutron are to be obtained by exchanging the values of the up and down quarks.

When using PDFs at a fixed factorization scale, we may need to
evaluate the radiative correction between this scale and the UV scale.
The anomalous dimensions of spin-2 twist-2 operators at the two-loop
level are given by \cite{Floratos:1978ny}
\begin{eqnarray}
 \mu \frac{d}{d\mu}(C^q_{{T}_i}, C^G_{{T}_i})=
(C^q_{{T}_i}, C^G_{{T}_i})~ \Gamma_{\rm T},
\end{eqnarray}
with $\Gamma_{\rm T}$ an $(N_f+1)\times (N_f+1)$ matrix:
\begin{eqnarray}
 \Gamma_{\rm T}&=&
\left(
\begin{tabular}{ccccc}
$\gamma_{qq}$ & $0$ & $\cdots$ & $0$ & $\gamma_{qg}$ \\
$0$     &$\gamma_{qq}$ & &$\vdots$ & $\vdots$\\
$\vdots$ & & $\ddots$ & $0$ & $\vdots$ \\
$0$ & $\cdots$ & $0$ & $\gamma_{qq}$ & $\gamma_{qg}$\\
$\gamma_{gq}$ & $\cdots$ & $\cdots$ & $\gamma_{gq}$ & $\gamma_{gg}$
\end{tabular}
\right)
,
\end{eqnarray}
where
\begin{eqnarray}
 \gamma_{qq}&=&\frac{16}{3}C_F\cdot\frac{\alpha_s}{4\pi}
+\biggl(-\frac{208}{27}C_FN_f-\frac{224}{27}C_F^2+\frac{752}{27}C_FN_c\biggr) 
\biggl(\frac{\alpha_s}{4\pi}\biggr)^2~,\nonumber \\
 \gamma_{qg}&=&\frac{4}{3}\cdot\frac{\alpha_s}{4\pi}
+\biggl(\frac{148}{27}C_F+\frac{70}{27}N_c\biggr) 
\biggl(\frac{\alpha_s}{4\pi}\biggr)^2~,\nonumber \\
 \gamma_{gq}&=&\frac{16}{3}C_F\cdot\frac{\alpha_s}{4\pi}
+\biggl(-\frac{208}{27}C_FN_f-\frac{224}{27}C_F^2+\frac{752}{27}C_FN_c\biggr) 
\biggl(\frac{\alpha_s}{4\pi}\biggr)^2~,\nonumber \\
 \gamma_{gg}&=&\frac{4}{3}N_f\cdot\frac{\alpha_s}{4\pi}
+\biggl(\frac{148}{27}C_FN_f+\frac{70}{27}N_cN_f\biggr) 
\biggl(\frac{\alpha_s}{4\pi}\biggr)^2~.
\end{eqnarray}

\subsection{Matrix Elements and RG Equations for Axial vector operators}

The last one is the matrix elements of axial vector currents, 
\begin{eqnarray}
\langle N|\bar{q}\gamma_\mu\gamma_5 q|N\rangle
&\equiv&
2 S_\mu\Delta q_N,
\end{eqnarray}
where $S_\mu$ is for the nucleon spin and $\Delta q_N$ is the spin fraction of quark $q$. The spin fractions of the quark are measured in DIS to be \cite{Adams:1995ufa}
\begin{eqnarray}
\Delta u_p=0.77,\nonumber\\
\Delta d_p=-0.47,\nonumber\\
\Delta s_p=-0.15.
\end{eqnarray}
 Those of the neutron are to be obtained by exchanging the values of the up and down quarks. The axial vector currents are RG invariant at the one-loop level.

\subsection{Effective Interactions of WIMPs with Nucleons}

The effective interactions of WIMPs with nucleons are evaluated
using the nucleon matrix elements given in the previous subsection,
as
\begin{eqnarray}
 f_N/m_N&=& \sum_{q=u,d,s}C_s^q(\mu_H) f_{T_q}^{(N)}\nonumber\\
 &&+C_S^g(\mu_H)\frac{4\alpha_s^2(\mu_H)}{\pi\beta(\alpha_s(\mu_H))}
 \left(1-(1-\gamma_m(\mu_H))\sum_{q=u,d,s} f_{T_q}^{(N)}\right)
 \nonumber\\
 &&+ \frac34 \sum_{i=1,2}\sum_q^{N_f} C_{Ti}^q(\mu)(q(2;\mu)+\bar{q}(2;\mu))
 -\frac34 \sum_{i=1,2}C_{Ti}^g(\mu)g(2;\mu),
\label{spinidcoupling}\\
 a_N &=& \sum_{q=u,d,s}C_{AV}^q(\mu_H)\Delta q.
\end{eqnarray}
As mentioned above, $C_S^q$, $C_S^g$, and $C_{AV}^q$ are RG-invariant at the one-loop
level. When calculating $f_N$ and $a_N$ at the leading order of $\alpha_s$,
\begin{eqnarray}
 C_S^q(\mu_H)&=&C_S^q(\mu_{UV}),~~(q=u,d,s),\nonumber\\
 C_S^g(\mu_H)&=&C_S^g(\mu_{UV})-\frac1{12} \sum_{q=c,b,t} C_S^q(\mu_{UV}),\nonumber\\
 C_{AV}^q(\mu_H) &=& C_{AV}^q(\mu_{UV}),~~(q=u,d,s).
\end{eqnarray}
Thus, the scalar operator contribution to $f_N$ is simplified as 
\begin{eqnarray}
 f_N/m_N|{\rm scalar~op.}&=& \sum_{q=u,d,s}C_S^q(\mu_{UV}) f_{T_q}^{(N)}\nonumber\\
 &&-\frac89 (C_S^g(\mu_{UV})-\frac1{12} \sum_{q=c,b,t} C_S^q(\mu_{UV}))
 \left(1-\sum_{q=u,d,s} f_{T_q}^{(N)}\right).\nonumber\\
&&
\end{eqnarray}

\section{Examples (at Leading-Order of $\alpha_s$)}
\label{sec:5}

We will now evaluate the effective interaction of WIMPs with the nucleon at the
leading order of $\alpha_s$, using the formulae in the previous section.
We consider three models: 1) gauge-singlet WIMPs coupled with the
Higgs boson, 2) gauge-singlet WIMPs coupled with colored scalars and
quarks, and 3) SU(2)$_L$ non-singlet WIMPs. When the effective
couplings of the WIMPs at the parton level are evaluated by integrating out
heavy particles at the UV scale, we have to consider the case of matching the
Wilson coefficients between the UV and effective theories.

\subsection{ Gauge Singlet WIMPs Coupled with Higgs Boson}

We now consider a case where the WIMPs are $SU(2)_L$ singlet fermions $X$, coupled with the SM Higgs boson $h$ as
\begin{eqnarray}
L_{\rm int}=-f_X\bar{X}X h.
\label{WIMP-Higgs}
\end{eqnarray}
This interaction is not symmetric under
SU(2)$_L\times$U(1)$_Y$. However, this interaction is introduced in
models where an SU(2)$_L$ singlet Higgs boson, coupled with $X$, is
introduced, and it is mixed with the SM Higgs boson. Alternatively, such as in the
SUSY SM, SU(2)$_L$ singlet and doublet fermions couple
with the SM Higgs boson so that Eq.~\ref{WIMP-Higgs} is generated due
to mixing of those fermions.

\begin{figure}
\begin{center}
 \begin{picture}(400,100)(0,0)
  \ArrowLine(130,50)(150,80)
  \ArrowLine(150,20)(130,50)
  \Text(155,20)[]{$q$}
  \Line(50,80)(70,50)
  \Line(70,50)(50,20)
  \Text(45,20)[b]{$X$}
  \DashLine(130,50)(70,50){2}
  \Text(100,55)[b]{$h$}
  \Text(165,90)[b]{(a)}  

  \ArrowLine(315,50)(335,30)
  \ArrowLine(335,30)(335,70)
  \ArrowLine(335,70)(315,50)
  \Gluon(335,70)(345,85){3}{2}
  \Gluon(335,30)(345,15){3}{2}
  \Text(320,68)[]{$q$}
  \Line(250,80)(270,50)
  \Line(270,50)(250,20)
  \Text(245,20)[b]{$X$}
  \DashLine(315,50)(270,50){2}
  \Text(292,55)[b]{$h$}
  \Text(360,90)[b]{(b)}  
\end{picture}
\end{center}
\caption{ Diagrams for the effective couplings of singlet WIMPs induced by
 Higgs boson exchange at the parton level. }
\label{fig2}
\end{figure}
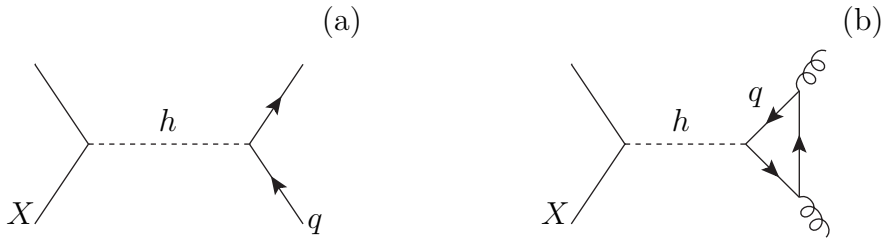

After integrating out the Higgs boson (Fig.~\ref{fig2}), the quark scalar operators are
generated as
\begin{eqnarray}
C_S^q(\mu_{UV})&=&\frac{1}{v_h m_h^2} f_X,
\end{eqnarray}
where $v_h$ is the vacuum expectation value of the Higgs field
$(v_H\simeq 246$~GeV) and $m_h$ is the SM Higgs mass ($m_h\simeq
125$~GeV). The gluon scalar operator is generated by the integration of the
heavy quarks, and the other operators are not generated at the leading
order of $\alpha_s$. Thus, the SI coupling constants with nucleons are given at
the leading order of $\alpha_s$ by%
\begin{eqnarray}
f_N/m_N&=&\frac{1}{v_h m_h^2} f_X \left(\bar{f}_{T_q}^{(N)}+3\times \frac2{27}
(1-\bar{f}_{T_q}^{(N)})\right),
\end{eqnarray}
where $\bar{f}_{T_q}^{(N)}\equiv\sum_{q=u,d,s}f_{T_q}^{(N)}$. From
this result, the SI cross section of proton is
$\sigma_{\rm SI}^p\simeq 2\times 10^{-42}{\rm cm}^2 \times f_X^2$. The
upperbound on $\sigma_{\rm SI}^p$ derived by XENON1T is about
$10^{-46}{\rm cm^2}$ for $M_X\simeq 50~{\rm GeV}$. Thus, $f_X\lsim
10^{-2}$ for $M_X\simeq 50~{\rm GeV}$.

\subsection{Gauge Singlet WIMPs
Coupled with Colored Scalars and Quarks}

Next, we consider the case where colored scalars are introduced and the
SU(2)$_L$ singlet WIMPs couple with the quarks and the colored
scalars. Binos in the SUSY SM have such couplings with scalar quarks.
Thus, this example corresponds to the limit of a heavy Higgsino 
in the SUSY SM, where the Bino--Higgs coupling is suppressed.
 
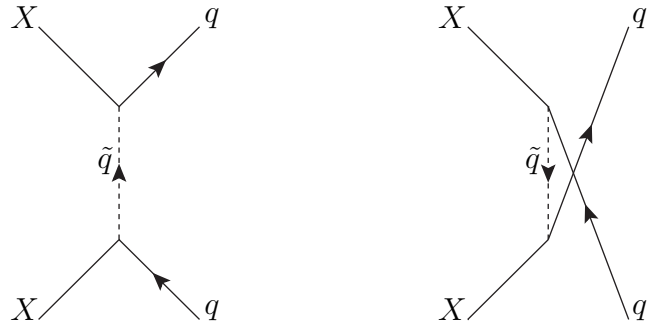
\begin{figure}
\begin{center}
 \begin{picture}(300,150)(0,0)
  \DashArrowLine(70,50)(70,100){2}
  \Text(65,75)[b]{$\tilde{q}$}
  \Line(70,50)(40,20)
  \Text(35,20)[b]{$X$}
  \ArrowLine(100,20)(70,50)
  \Text(105,20)[b]{$q$}
  \Line(70,100)(40,130)
  \Text(35,130)[b]{$X$}
  \ArrowLine(70,100)(100,130)
  \Text(105,130)[b]{$q$}
  \DashArrowLine(230,100)(230,50){2}
  \Text(225,75)[b]{$\tilde{q}$}
  \ArrowLine(260,20)(230,100)
  \Text(265,20)[b]{$q$}
  \Line(200,20)(230,50)
  \Text(195,20)[b]{$X$}
  \ArrowLine(230,50)(260,130)
  \Text(265,130)[b]{$q$}
  \Line(230,100)(200,130)
  \Text(195,130)[b]{$X$}
 \end{picture}
\end{center}
\caption{
 Diagrams for the effective couplings of singlet WIMPs with quarks induced by colored scalar exchange. }
\label{fig3}
\end{figure}

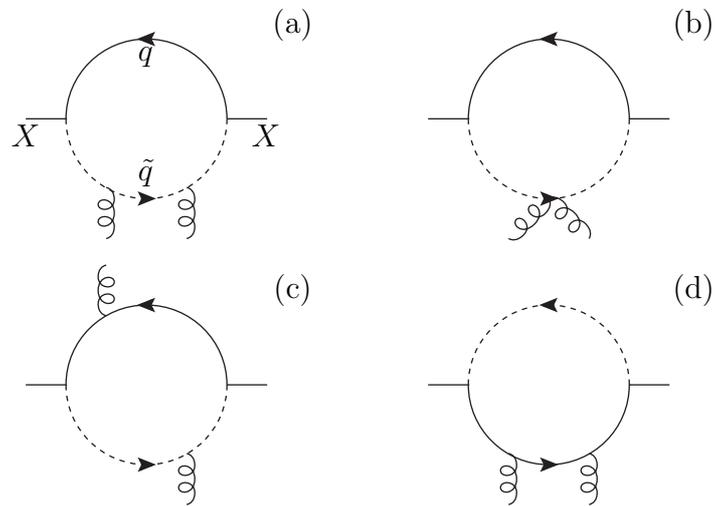
\begin{figure}
\begin{center}
 \begin{picture}(300,200)(0,0)
  \ArrowArc(75,50)(30,0,180)
  \DashArrowArc(75,50)(30,180,360){2}
  \Line(105,50)(120,50)
  \Line(45,50)(30,50)
  \Gluon(60,75.980)(60,95){3}{2}
  \Gluon(90,24.192)(90,5){3}{2}
  \Text(130,80)[b]{(c)}
  \DashArrowArc(225,50)(30,0,180){2}
  \ArrowArc(225,50)(30,180,360)
  \Line(255,50)(270,50)
  \Line(195,50)(180,50)
  \Gluon(210,24.192)(210,5){3}{2}
  \Gluon(240,24.192)(240,5){3}{2}
  \Text(280,80)[b]{(d)}
  \ArrowArc(75,150)(30,0,180)
  \Text(75,170)[b]{$q$}
  \DashArrowArc(75,150)(30,180,360){2}
  \Text(75,125)[b]{$\tilde{q}$}
  \Line(105,150)(120,150)
  \Text(120,140)[b]{$X$}
  \Line(45,150)(30,150)
  \Text(30,140)[b]{$X$}
  \Gluon(60,124.192)(60,105){3}{2}
  \Gluon(90,124.192)(90,105){3}{2}
  \Text(130,180)[b]{(a)}
  \ArrowArc(225,150)(30,0,180)
  \DashArrowArc(225,150)(30,180,360){2}
  \Line(255,150)(270,150)
  \Line(195,150)(180,150)
  \Gluon(225,120)(210,105){3}{2}
  \Gluon(225,120)(240,105){3}{2}
  \Text(280,180)[b]{(b)}
 \end{picture}
\end{center}
\caption{ Diagrams for the effective couplings of singlet WIMPs with gluons induced by colored scalar exchange. }
\label{fig4}
\end{figure}

The interactions of WIMPs with quarks and colored scalars are given
by 
\begin{eqnarray}
L_{\rm int}&=& \sum_q \bar{q}(a_q+b_q\gamma_5)X\tilde{q}+{\rm h.c.},
\end{eqnarray}
where $\tilde{q}$ is the colored scalar. The $t$-channel colored scalar exchange
diagrams (Fig.~\ref{fig3}) generate the quark scalar, twist-2, and axial vector
operators, while the gluon scalar and twist-2 operators come from
one-loop diagrams. Now, consider the leading contribution of
$\alpha_s$ to the effective couplings of the WIMPs with nucleons. Thus,
the one-loop contribution to the gluon scalar operator ((Fig.~\ref{fig4})) is included in
the evaluation, while that to the gluon twist-2 operator is subleading,
and is thus neglected.

From direct calculation, the Wilson coefficients for the quark operators
are derived as 
\begin{eqnarray}
C_S^{q}(\mu_{UV})&=&\frac{a_q^2-b_q^2}{4 m_q}\frac{1}{M_X^2-M_{\tilde{q}}^2}
+\frac{a^2_q+b^2_q}8 \frac{M_X}{(M_X^2-M_{\tilde{q}}^2)^2},\nonumber\\
C_{T1}^q(\mu_{UV})&=&\frac{a_q^2+b^2_q}2\frac{M_X}{(M_X^2-M_{\tilde{q}}^2)},\nonumber\\
C_{T2}^q(\mu_{UV})&=&0,\nonumber\\
C_{AV}^q(\mu_{UV})&=&-\frac{a_q^2+b_q^2}4 \frac{1}{M_X^2-M_{\tilde{q}}^2},
\end{eqnarray}
where $M_{\tilde{q}}$ is the colored scalar mass and
$\mu_{UV}\simeq M_{\tilde{q}}$. The above Wilson coefficients are for quarks $q$
with $m_q<\mu_{UV}$. 

On the other hand, we have to match the UV and effective theories in
order to derive the Wilson coefficient for the gluon scalar
operator. The evaluation of the Wilson coefficients for operators
including gluon field strengths is always troublesome due to the
tensor structure. However, when the momenta of the external gluons are negligibly small
and the gluon fields may be included as background fields, the
Fock--Schwinger gauge ($x^\mu A^A_\mu (x)=0$) is quite convenient for
evaluating them, since we introduce propagators under the gluon field
strength background with the gauge. Details of the technique are given in
Appendix~\ref{FSgauge}.

Four one-loop diagrams in Fig.~\ref{fig4} contribute to the
gluon scalar operator. While Diagrams $A$ and $C$ are automatically zero
in the Fock-Schwinger gauge, Diagrams $B$ and $D$ give contributions to the gluon scalar operator as
\begin{eqnarray}
C_S^g|_B&=&\sum_q\left[
\frac{a_q^2+b_q^2}{16}\frac{M_X}{6 M_{\tilde{q}}^2(M_X^2-M_{\tilde{q}}^2)}
\right],\nonumber\\
C_S^g|_D&=&\sum_q\left[
-\frac{a_q^2+b_q^2}{16}\frac{M_X}{6 (M_X^2-M_{\tilde{q}}^2)^2}
-
\frac{a_q^2-b_q^2}{16}\frac{1}{
3m_q M_{\tilde{q}}^2(M_X^2-M_{\tilde{q}}^2)}
\right].
\end{eqnarray}
Here, we take the leading terms of $m_q$ assuming $m_q\ll
M_{\tilde{q}}$ to demonstrate the matching. We find that
$C_S^g|_D=-1/12 \sum_q C_S^q$. The contribution of Diagram $D$
corresponds to the integration of heavy quarks in the effective
theory. Thus,
\begin{eqnarray}
C_S^g(\mu_{UV})
&=&\sum_{q={\rm all}} C_S^g|_D+
\sum_{q(m_q>\mu_{UV})} C_S^g|_D.
\end{eqnarray}
In the above equations, the leading term of $m_q$ is shown, though we should 
use the exact formulae when $m_q$ is not negligible compared to 
$M_X$ and $M_{\tilde{q}}$.

We can now show some numerical results. We assume that the singlet WIMPs
have interactions with only the top and bottom quarks ($a_q=b_q=0$ for
$q=u,d,c,s$ and $a_q=b_q=1/2$ for $q=t,b$). This is a similar to the
case where the binos are coupled with only the third-generation quarks and
scalar quarks in the SUSY SM, assuming the other squarks are decoupled.

First, we should discuss the renormalization scale-dependence of the contribution
from the twist-2 operators to the SI coupling constants of WIMPs with
nucleons (the last two terms in Eq.~\ref{spinidcoupling}). The Wilson
coefficients for the quark/gluon scalar operators are RG invariant at
the leading order of $\alpha_s$, while those for the twist-2 operators
are scale-dependent. As in Eq.~\ref{spinidcoupling}, they may be
evaluated at any scale if the factorization scale of the PDFs is properly
chosen.

\begin{figure}
\begin{center}
\includegraphics[width=6cm]{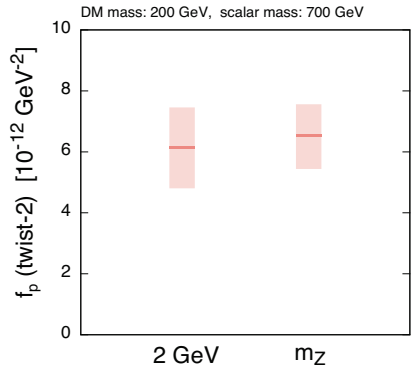}
\end{center}
\caption{
Contribution from twist-2
operators to the SI coupling constant of WIMPs with protons,
evaluated at $\mu=2~$GeV and $m_Z$. An explanation for the figure is given in the text.
The figure comes from Ref.~\cite{Hisano:2015bma}.
\label{fig5}}
\end{figure}

In the Fig.~\ref{fig5} the contribution from the twist-2
operators to the SI coupling constants of WIMPs with protons is shown, which is
evaluated at $\mu=2~$GeV and $m_Z$. The PDFs for $\mu=2~$GeV and $m_Z$
are also used. Here, we take $M_X=200~$GeV and
$M_{\tilde{q}}=700~$GeV. The red and light-pink bars denote the
uncertainties coming from the PDF input and the perturbation in
$\alpha_s$, respectively. The method for evaluating the uncertainty
from the PDFs' inputs is described in Ref.~\cite{Owens:2012bv}. The
uncertainty caused by neglecting the higher-order contributions in
$\alpha_s$ is evaluated by varying the input and quark-mass threshold
scales by a factor of two ($M_{\tilde{q}}/2 \leq \mu \leq 2
M_{\tilde{q}}$, $m_t/2 \leq\mu \leq 2m_t$, and so on). 
The center values of the two calculations are almost the same, and the uncertainty from the perturbation in $\alpha_s$ for the
case $\mu=2~$GeV is slightly larger. This is expected from the
nature of asymptotic freedom in QCD. Thus, it is better to evaluate
the contribution from the twist-2 operators at the weak scale.

\begin{figure}
\begin{center}
 \includegraphics[width=6cm]{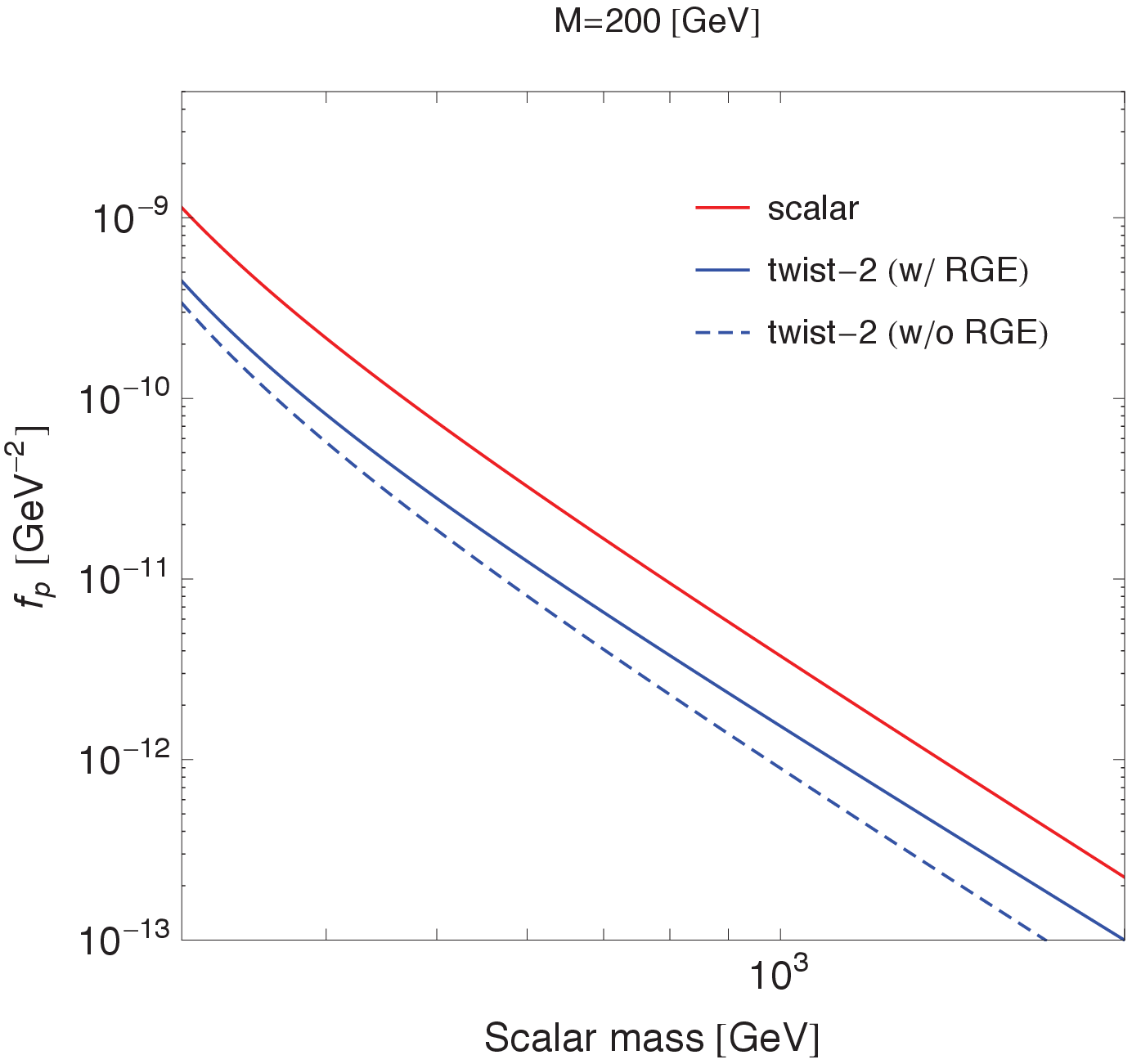}
 \includegraphics[width=6cm]{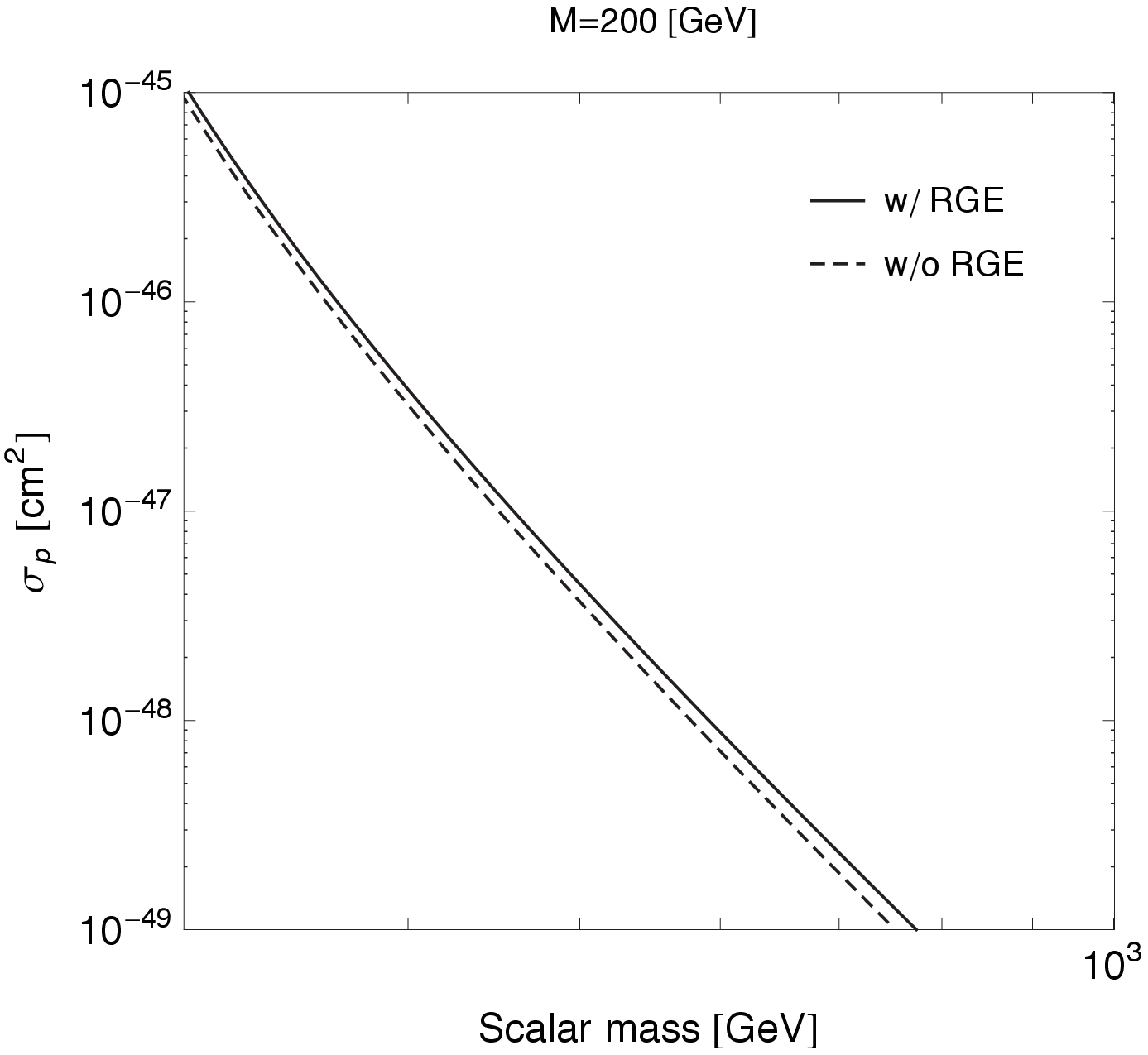}
\end{center}
\caption{
SI coupling constant (left) and SI
cross section of WIMPs with protons (right) as functions of colored scalar
mass. Here, $M_X=200~$GeV. Explanation for figures is given in text.
Figures come from Ref.~\cite{Hisano:2015bma}.
\label{fig6}}
\end{figure}

Next, the SI coupling constant (left) and the SI cross section
of WIMPs with protons (right) as functions of the colored scalar mass are shown
in Fig.~\ref{fig6}. Here, $M_X=200~$GeV again. In the
left-hand figure, the upper (lower) solid line shows the
contribution of the scalar-type (twist-2-type) operators to the SI
coupling constant of WIMPs with protons. For the twist-2 contribution,
we use PDFs at $\mu=m_Z$ and show the calculations both with and without
the RG effects between $M_{\tilde{q}}$ and $m_Z$ in the solid and dashed
lines, respectively. In the right-hand figure, the SI cross section
of WIMPs with protons is shown with (solid) and without the RG
effects (dashed line). The RG effects change the
resulting value for the twist-2 contribution to the SI coupling constant by
more than 50$\%$ when $M_{\tilde{q}}=500~$GeV, and the scattering cross
sections are modified by more than 20$\%$. Thus, if the colored
mediator mass is much larger than the factorization scale in the PDFs
adopted in the evaluation, it is important to include the
RG effects.

\subsection{SU(2)$_L$ non-singlet WIMPs}

Neutral components in the SU(2)$_L$ multiplets are candidates for WIMPs.
If they are Dirac fermions or complex scalars with nonzero hypercharges, 
the DM direct detection experiments impose a strict bound on them. If the WIMP ($X$)
is a fermion with hypercharge $Y_X$, the neutral current interaction induces
SI interactions with nucleons ($N=n,p$) as follows,
\begin{eqnarray}
{L}_{\rm eff}&=&
\sqrt{2} G_FY_X \bar{X} \gamma_\mu X\left[(1-4 \sin^2\theta_W) \bar{p}\gamma_\mu p-
\bar{n}\gamma_\mu n\right]
\end{eqnarray}
where $G_F$ and $\theta_W$ are the Fermi constant and Weinberg angle,
respectively. The SI cross section of WIMPs with neutrons is larger
than that with protons due to the accidental cancellation in the
latter. They are insensitive to the WIMP mass, and the SI cross
section with neutrons is approximately given by $Y_X^2\times 7\times
10^{-40}$cm$^2$. If the WIMPs are the dominant component of the DM in
our galaxy, the mass should be larger than $\sim 10^5~$TeV. This is much
heavier than the unitarity bound on the WIMP mass in the assumption
that the WIMPs are in thermal equilibrium in the early universe
\cite{Griest:1989wd}. Even if the WIMPs are complex scalars with nonzero
hypercharge , a similar bound is derived. On the other hand,
this constraint is avoidable if the WIMPs are Majorana fermions or
real scalars, since their vector coupling is forbidden automatically.

SU(2)$_L$ triplet Majorana fermions with zero hypercharge are one of
the examples. They are called winos in the SUSY SM, and they are
superpartners of the weak gauge bosons. The triplet fermions include a
neutral Majorana fermion and a charged Dirac fermion. The radiative
corrections due to the SU(2)$_L$ gauge interactions generate the mass
splitting between the neutral and charged fermions as $\Delta
M_X\simeq 165$~MeV when $M_X\gg m_Z$ \cite{Ibe:2012sx}, and it makes
the neutral fermion lighter than the charged one. The radiative
corrections may easily dominate over the contribution to the mass
splitting from effective operators induced by the integration of heavier
particles, since the operators have mass dimensions of seven and above.

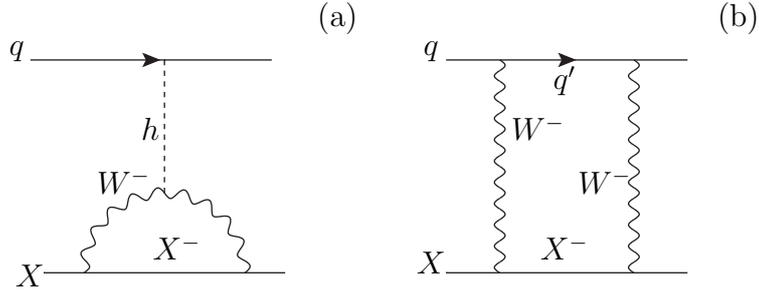
\begin{figure}
\begin{center}
 \begin{picture}(300,130)(0,0)
  \ArrowLine(25,110)(115,110)
  \Text(20,110)[b]{$q$}
  \Line(30,30)(120,30)
  \Text(25,25)[b]{$X$}
  \DashLine(75,110)(75,60){2}
  \Text(70,80)[b]{$h$}
  \PhotonArc(75,30)(30,0,180){2}{10}
  \Text(60,60)[b]{$W^-$}  
  \Text(80,35)[b]{$X^-$}
  \Text(140,120)[b]{(a)}  
  \Text(175,110)[b]{$q$}
  \ArrowLine(180,110)(270,110)
  \Text(225,97)[b]{$q'$}
  \Text(175,30)[b]{$X$}
  \Line(180,30)(270,30)  
  \Text(215,80)[b]{$W^-$}
  \Text(240,60)[b]{$W^-$}
  \Photon(200,110)(200,30){2}{10}
  \Photon(250,110)(250,30){2}{10}
  \Text(225,35)[b]{$X^-$}
  \Text(290,120)[b]{(b)}  
 \end{picture}
\end{center}
\caption{ Diagrams for effective couplings of SU(2)$_L$ triplet fermion WIMPs with quarks. }
\label{winodiagram1}
\end{figure}

\begin{figure}
\begin{center}
 \begin{picture}(150,150)(0,0)
  \Text(52,110)[b]{$q$}
  \Line(30,30)(120,30)
  \Text(25,25)[b]{$X$}
  \DashLine(75,100)(75,60){2}
  \ArrowLine(75,100)(50,125)
  \ArrowLine(50,125)(100,125)
  \ArrowLine(100,125)(75,100)
  \Gluon(50,125)(35,140){2}{2}
  \Gluon(100,125)(115,140){2}{2}
  \Text(70,80)[b]{$h$}
  \PhotonArc(75,30)(30,0,180){2}{10}
  \Text(60,60)[b]{$W^-$}  
  \Text(80,35)[b]{$X^-$}
  \Text(140,145)[b]{(a)}  
 \end{picture}

  \begin{picture}(450,150)(150,0)
  \Text(225,105)[b]{$q$}
  \Text(225,45)[b]{$q'$}
  \Line(180,30)(270,30)  
  \Text(200,45)[b]{$W^-$}
  \Text(260,45)[b]{$W^-$}
  \ArrowArc(225,85)(30,0,360)
  \Photon(200,30)(210,60){2}{4}
  \Photon(250,30)(240,60){2}{4}
  \Gluon(200,103)(185,140){2}{4}
  \Gluon(250,103)(265,140){2}{4}
  \Text(290,145)[b]{(b)}  
  %
  \Line(330,30)(420,30)  
  \Text(350,45)[b]{$W^-$}
  \Text(410,45)[b]{$W^-$}
  \Text(375,105)[b]{$q$}
  \Text(375,45)[b]{$q'$}
  \ArrowArc(375,85)(30,0,360)
  \Photon(350,30)(360,60){2}{4}
  \Photon(400,30)(390,60){2}{4}
  \Gluon(375,55)(335,140){2}{10}
  \Gluon(400,103)(415,140){2}{4}
  \Text(440,145)[b]{(c)}  
  %
 \end{picture}

\end{center}
\caption{Diagrams for the effective couplings of SU(2)$_L$ triplet fermion WIMPs with gluons}
\label{winodiagram2}
\end{figure}
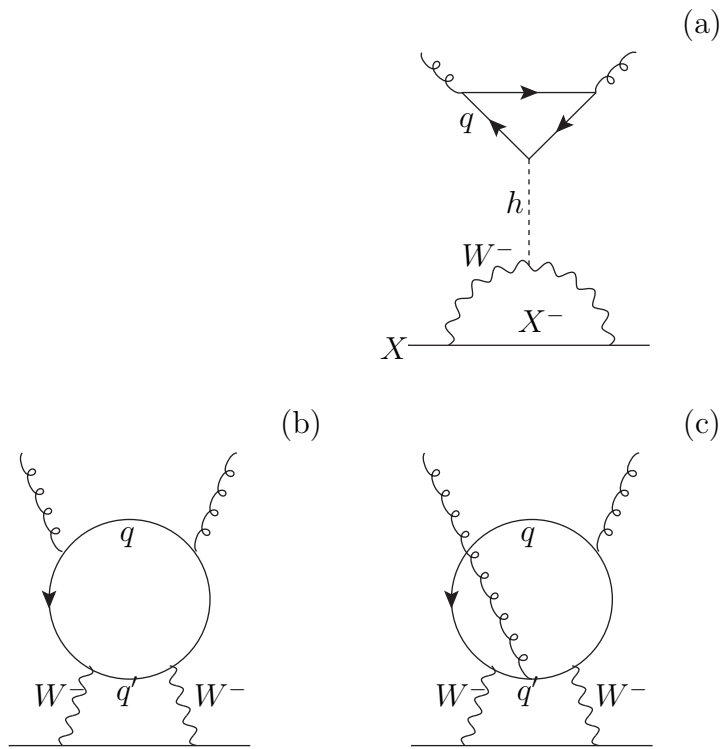

Now, assume that the triplet fermions have only gauge
interactions.\footnote{
In the SUSY SM, this situation is realized when the superpartners, except for gauginos, are decoupled such as in \cite{Giudice:1998xp}. 
}
 The WIMPs (the neutral component, $X$) 
only interact with the $W$ boson and the charged component $X^-$:
\begin{eqnarray}
L_{int}=-g_2(\bar{X}\gamma^\mu X^- W_{\mu}^{\dagger}+{\rm h.c.}).
\end{eqnarray}
In this case, the WIMPs couple with quarks at the one-loop level
(Fig.~\ref{winodiagram1}), and with gluons at the two-loop level
(Fig.~\ref{winodiagram2}). The diagrams give finite contributions,
and the loop momenta in the diagrams are typically around
$m_W$. Thus, we consider the effective theory with $N_f=5$ flavors at
$\mu_{UV}=m_Z$.

 The Higgs boson exchange diagrams in Fig.~\ref{winodiagram1}
(a) and Fig.~\ref{winodiagram2} (a) generate the scalar operators
\begin{eqnarray}
C_S^q(\mu_{UV})|_{\rm Higgs}&=&
\frac{\alpha^2_2}{4m_Wm_h^2}g_H(\omega),~~(q=u,d,s,c,b),\nonumber\\
C_S^g(\mu_{UV})|_{\rm Higgs}&=&
-\frac{\alpha_2^2}{48m_Wm_h^2}g_{H}(\omega)
\end{eqnarray}
where $\omega=m_W^2/M_X^2$ and $\alpha_2=g_2^2/4\pi$.
The coefficient $C_S^g(\mu_{UV})$ comes from the integration of the top quark.
Here and in the following calculations, the mass difference between the charged
and neutral fermions is neglected.
The mass functions including $g_H(\omega)$ in this section are given in 
Ref.~\cite{Hisano:2015rsa}.

Next, let us calculate the contributions to the scalar operators from
the box diagrams in Fig.~\ref{winodiagram1} (b) and
Fig.~\ref{winodiagram2} (b) and (c). In this calculation, it is
convenient to first derive the OPEs of the charged current-charged current
correlator, 
\begin{eqnarray}
\Pi^W_{\mu\nu}(q)\equiv i\int d^4 x ~{\rm e}^{iqx}
T\left\{
J_{\mu}^W(x)J^W_\nu(0)^\dagger
\right\},
\end{eqnarray}
where 
\begin{eqnarray}
J^W_\mu(x)=\sum_{i=1,2,3}\frac{g_2}{\sqrt{2}}\bar{u}_i\gamma_\mu P_L d_i. 
\end{eqnarray}
The correlator is decomposed into its transverse $\Pi^W_T(q^2)$ and 
longitudinal parts $\Pi^W_L(q^2)$,
\begin{eqnarray}
\Pi^W_{\mu\nu}(q)&=&(
(-g_{\mu\nu} +\frac{q_\mu q_\nu}{q^2})
\Pi^W_T(q^2)
 +\frac{q_\mu q_\nu}{q^2}
\Pi^W_L(q^2).
\end{eqnarray}
When connecting the correlator to the WIMP lines as in
Fig.~\ref{winodiagram1} (b) and Fig.~\ref{winodiagram2} (b) and (c),
the longitudinal part does not contribute to the scalar operators due
to the gauge invariance \cite{Hisano:2010fy}.

The quark and gluon scalar operators in the transverse
part of the correlator are represented by 
\begin{eqnarray}
\Pi^W_{T}(q)|_{\rm scalar}&=&
\sum_q^{N_f=5} c_{W,S}^q(q^2)m_q \bar{q}q
+c_{W,S}^g(q^2)\frac{\alpha_s}{\pi} G_{\mu\nu}^A G^{A\mu\nu}.
\end{eqnarray}
The tree-level diagrams contribute to $c_{W,S}^q(q^2)$, though the 
$c_{W,S}^q(q^2)$ are suppressed by the tiny quark masses,
since $J^W_\mu(x)$ is the $V$-$A$ current. The exception is 
 $c_{W,S}^b(q^2)$,
\begin{eqnarray}
 c^b_{W,S}(q^2) &=&\frac{g_2^2m_t^2}{8(q^2-m_t^2)^2},
\end{eqnarray}
due to the large top quark mass. On the other hand,
the one-loop diagrams induce $c_{W,S}^g(q^2)$ as 
\begin{eqnarray}
c_{W,S}^g(q^2)&=& \frac{g^2_2}{48 q^2}\left[2 + \frac{q^2}{q^2-m_t^2}\right],
\end{eqnarray}
where the first and second terms in the bracket come from loop diagrams including the first two
and third generations, respectively. Using these results, 
the Wilson coefficients for the scalar operator are
\begin{eqnarray}
C_S^b(\mu_{UV})|_{\rm box}&=&\frac{\alpha_2^2}{m_W^3} \left[(-3)g_{\rm btm}(\omega,\tau)\right],
\nonumber\\
C_S^g(\mu_{UV})|_{\rm box}&=&
\frac{\alpha_2^2}{4m_W^3} \left[2g_{\rm B1}(\omega)+g_{\rm top}(\omega,\tau)\right]
\end{eqnarray}
where $\tau=m_t^2/M_X^2$.

The contributions to the twist-2 operators also come from 
Fig.~\ref{winodiagram1} (b) and Fig.~\ref{winodiagram2} (b) and (c),
though those to the gluon twist-2 operators are at $O(\alpha_s)$ so that 
they are neglected here. The OPEs in the charged current-charged current correlator involve the
quark twist-2 operators at the leading order of $\alpha_s$: 
\begin{eqnarray}
\Pi^W_{\mu\nu}
&=&
\sum_{q=u,d,c,s}
\frac{g_2^2}2
\left[
-\frac{g_{\mu\rho}g_{\nu\sigma}q^2-g_{\mu\rho}q_\nu q_\sigma-q_\mu q_\rho g_{\nu\sigma}
+g_{\mu\nu}q_{\rho}q_{\sigma}}{(q^2)^2} 
\right]O^{q\rho\sigma}
\nonumber\\
&&
+\frac{g_2^2}2
\left[-\frac{g_{\mu\rho}g_{\nu\sigma}(q^2-m_t^2)-g_{\mu\rho}q_\nu q_\sigma-q_\mu q_\rho g_{\nu\sigma}+g_{\mu\nu}q_{\rho}q_{\sigma}}{(q^2-m_t^2)^2}
\right] O^{b\rho\sigma}.
\end{eqnarray}
Then, the Wilson coefficients for the quark twist-2 operators are 
given by
\begin{eqnarray}
C_{Ti}^q(\mu_{UV})&=&\frac{\alpha_2^2}{m_W^3}g_{Ti}(\omega,0),\nonumber\\
C_{Ti}^b(\mu_{UV})&=&\frac{\alpha_2^2}{m_W^3}g_{Ti}(\omega,\tau),
\end{eqnarray}
($i=1,2$).

The axial vector coupling is also evaluated in a similar way: 
\begin{eqnarray}
C^q_{AV}&=&\frac{\alpha_2^2}{8m_W^2} g_{AV}(\omega).
\end{eqnarray}

We have now shown the Wilson coefficients at the leading order of
$\alpha_s$. Some of the Wilson coefficients are not
suppressed even if $M_X$ is much heavier than $m_W$. The Wilson
coefficients for the gluon and scalar operators at the hadronic
scale are
\begin{eqnarray}
C_S^q(\mu_H) &\simeq& \frac{\alpha_2^2}{4m_W m_h^2}(-2\pi),\nonumber\\
C_S^g(\mu_H) &\simeq& -3\frac{\alpha_2^2}{48m_W m_h^2}(-2\pi)
     +2\frac{\alpha_2^2}{4m_W^3}\frac{\pi}{12}
     +\frac{\alpha_2^2}{4m_W^3}\frac{\pi}{24}\frac{3x_{tw}+2}{(x_{tw}+1)^3},
\end{eqnarray}
for $M_X\gg m_t$, $m_W$. Here, $x_{tw}=m_t/m_W$. The first term in
$C_S^g(\mu_H)$ comes from the heavy quark contribution induced by
the Higgs exchange, and the second and third are from box diagrams
including the first two and third generations, respectively. The
last one is more suppressed by the top quark mass compared to
the box diagrams including light quarks.
The Wilson coefficients for the quark twist-2 operators at $\mu_{UV}$
are
\begin{eqnarray}
C_{T1}^q(\mu_{\rm UV}) &=&
\frac{\alpha_2^2}{m_W^3}\frac{\pi}3,\nonumber\\
C_{T2}^q(\mu_{\rm UV}) &=&
O(\frac{m_W}{M_X}).
\end{eqnarray}
The axial vector coupling constants are also suppressed by
$O({m_W}/{M_X})$. Thus, the SI coupling constants become independent of
$M_X$ in the limit of
$M_X\gg m_W$ while the SD coupling vanishes in the limit \cite{Hisano:2004pv,Hisano:2010fy}. This implies
that the SI cross sections are insensitive to the WIMP mass. This is
welcome in the search for the WIMPs. 

On the other hand, the contributions are comparable to each other so
that the accidental cancellation suppresses the SI cross sections. The
SI coupling constants of WIMPs with nucleons are roughly estimated as
\begin{equation}
 f_N/m_N\simeq
 \frac{\alpha_2^2}{m_W^3}\times 0.4
 -\frac{\alpha_2^2}{m_W^3}\times 0.27
 -\frac{\alpha_2^2}{m_W^3}\times 0.03,
\end{equation}
where the first, second, and third terms come from the quark twist-2
($C_{T1}^q(\mu_{\rm UV})$), gluon scalar ($C_S^g(\mu_H)$), and quark
scalar operators ($C_S^q(\mu_H)$), respectively. This cancellation reduces the SI
cross sections by a factor of more than 10. In addition, it decreases the
reliability of the calculation as the significance of the higher-order correction is relatively high \cite{Hill:2013hoa}. The
predicted SI cross sections are close to the neutrino SI scattering
background, the neutrino floor. Thus, a more reliable evaluation method for the
SI cross sections are needed.

The above calculation can be extended to the cases of SU(2)$_L$ $n$-plets \cite{Hisano:2011cs}. Even if the hypercharge is nonzero,
the WIMPs may be Majorana fermions or real scalars by introducing the
effective coupling with the Higgs boson
\cite{Hisano:2014kua}. Higgsinos in the SUSY SM are SU(2)$_L$ doublets
with hypercharges $\pm 1/2$, while the effective operator with the Higgs boson
induced by integrating out gauginos decomposes a neutral Dirac fermion
into two Majorana fermions with mass splitting.

\section{Towards the Next-Leading Order Calculation of $\alpha_s$}
\label{sec:6}

In the previous sections, we showed the evaluation of the elastic
scattering cross section of WIMPs with nucleons at the leading order
of $\alpha_s$. In many phenomenological studies, the leading-order
evaluation is enough, since the DM abundance around the earth still has
large uncertainties. However, in some cases, evaluation at
higher orders is required. 

First, as in the case of SU(2)$_L$ triplet fermions, the accidental
cancellation may suppress the leading order contribution. Many DM
direct experiments are more sensitive to the SI cross section than the
SD one. The SI cross section is induced from several contributions at the
parton level, which interfere with each other. In the case of
SU(2)$_L$ triplet fermions, the contributions to the SI coupling at the
parton level are comparable to each other. Then, the destructive
interference reduces the SI cross section significantly. The higher-order corrections have to be included in the evaluation of the SI
cross section in order to be reliable.

Next is the uncertainty from the UV scale, $\mu_{UV}$. The quark and
gluon scalar operators are RG invariant at the leading order of
$\alpha_s$.  Thus, their contribution is independent of $\mu_{UV}$ at
the leading order. On the other hand, the SI cross section sensitive
to the quark twist-2 operators at the leading order, and their Wilson
coefficients depend on $\mu_{UV}$. We need to include the next-leading
order contribution in order to reduce the uncertainty from $\mu_{UV}$.

Third, some next-leading order corrections are known to be large. For
example, it is known that the threshold correction to the gluon scalar
operator at the quark mass scale at the next-leading order is large.
 
The next-leading order evaluation of the elastic scattering cross
section of WIMPs with nucleons is not yet complete. Some 
results for the SI cross section are shown in this section. 

\subsection{Matching Condition at Quark Mass Threshold}

The Wilson coefficients for the quark and gluon scalar operators are
not RG invariant at $O(\alpha_s^2)$. Thus, the two-loop RG equations
for the Wilson coefficients need to be solved in order to derive
$C_S^q(\mu_H)$ and $C_S^g(\mu_H)$ with $\mu_H$ the hadronic scale at
the next-leading order of $\alpha_s$. When the factorization scale of the
PDFs, used in the evaluating the matrix elements of twist-2 operators,
is different from the UV scale, we also include the correction of
their two-loop RG equations. The RG equations for the Wilson
coefficients are shown in Section~\ref{sec:4}.

At the heavy quark mass thresholds, we also have to include the threshold
correction to the Wilson coefficients at $O(\alpha_s)$. When the
bottom quark is decoupled and $N_f$ is changed from 5 to 4, the
matching conditions for the Wilson coefficients at the decoupling
scale $\mu_b(\simeq m_b)$ are given by
\begin{eqnarray}
C_S^q(\mu_b)|_{N_f=4}
&=&
C_S^q(\mu_b)|_{N_f=5},~~(q=u,d,s,c),
\nonumber\\
\alpha_s(\mu_b)C_S^g(\mu_b)|_{N_f=4}
&=&
\left[1+\frac{\alpha_s(\mu_b)}{4\pi}\frac 23 \log\frac{m_b^2}{\mu_b^2}\right]
\alpha_s(\mu_b) C_S^g(\mu_b)|_{N_f=5}\nonumber\\
&&-\frac{\alpha_s(\mu_b)}{12}\left[1+
\frac{\alpha_s(\mu_b)}{4\pi}\left(11
+\frac 23 \log\frac{m_b^2}{\mu_b^2}\right)\right]
\alpha_s(\mu_b) C_S^b(\mu_b)|_{N_f=5},
\nonumber\\
C_{Ti}^q(\mu_b)|_{N_f=4}&=&C_{Ti}^q(\mu_b)|_{N_f=5},~~(i=1,2),
\nonumber\\
C_{Ti}^g(\mu_b)|_{N_f=4}&=&
\left[1+\frac{\alpha_s(\mu_b)}{4\pi}\frac 23 \log\frac{m_b^2}{\mu_b^2}\right]
C_{Ti}^g(\mu_b)|_{N_f=5}
\nonumber\\
&&
+\frac{\alpha_s(\mu_b)}{4\pi}\frac 23 \log\frac{m_b^2}{\mu_b^2}
~C_{Ti}^b(\mu_b)|_{N_f=5},,~~(i=1,2),
\end{eqnarray}
where
\begin{eqnarray}
 \frac{1}{\alpha_s(\mu_b)|_{N_f=4}}&=&\frac{1}{\alpha_s(\mu_b)|_{N_f=5}}+\frac{1}{(3\pi)}
\log\frac{\mu_b}{m_b}.
\end{eqnarray}
These matching conditions are derived by comparing the two effective
theories with $N_f=5$ and $4$. The logarithmic terms correspond to the
RG equations. The large factor of 11 in the matching condition for
$C_S^g$ is found in the second line of
Eq.~\ref{heavyquarkintegral}. Thus, the threshold correction $C_S^g$
at $\alpha_s$ is not negligible even if it is of the next-leading
order \cite{Djouadi:2000ck}.

Similar matching conditions are applied for the other heavy quark
threshold. For example, the matching conditions for the twist-2 operators
at $\mu_b$ are irrelevant to the practical calculation when the
factorization scale of the PDFs is higher than $\mu_b$. However, if
$\mu_{UV}$ is higher than the top quark mass and the factorization
scale is below the top quark mass, we have to include the radiative
corrections between the two scales, taking account the top quark
threshold. The matching conditions for $C_{Ti}^g$ have to be included
in the calculation.

\subsection{Wilson Coefficients at UV Scale and SI Cross Sections at
 the Next-Leading Order of $\alpha_s$}

We now evaluate the SI cross section at the next-leading order of
$\alpha_s$. The calculation is very involved. We have to evaluate the
Wilson coefficient $C_S^g(\mu_{UV})$ for the gluon scalar operator at
higher orders of $\alpha_s$ than for the other operators in the UV theory.
In the case of the Higgs portal singlet WIMPs, the calculation is the
straightforward. For singlet WIMPs coupled with colored
scalars and quarks, what we have to evaluate at the UV scale are the following:\\
\begin{center}
\begin{tabular}{|c||c|c|}
\hline
Operators &LO &NLO\\
\hline
Quark scalar & tree & 1 loop\\
Quark twist-2 & tree &1 loop\\
Gluon scalar & 1 loop & 2 loop\\
Gluon twist-2.& $-$ & 1 loop\\
\hline
\end{tabular}
\end{center}
~\\
The coefficient for the gluon twist-2 operator from the one-loop diagrams
is simultaneously calculated when calculating that of the gluon scalar
operator at the leading order. However, the other contributions at the
next-leading order have not yet been evaluated.

The last are the SU(2)$_L$ triplet fermions. When evaluating the SI cross
sections for the SU(2)$_L$ triplet fermions at the next-leading order of
$\alpha_s$, we have to calculate the two and three-loop diagrams. This calculation cannot be done by hand, if we include all the next-leading order
contributions. Fortunately, the quark and gluon scalar operators in the
charged current-charged current correlator $\Pi^W_{\mu\nu}$ are
evaluated at the three-loop level \cite{Broadhurst:1994qj}, and the twist-2
operators at the two-loop level \cite{Bardeen:1978yd}, assuming the quarks
are massless. In the leading order calculation, while the top quark
mass is not negligible, the contributions of the
third generation to the SI cross sections are suppressed by the top
quark mass itself. Thus, we can neglect the next-leading order contributions
of the third generation and evaluate the uncertainties from it.

In the Ref.\cite{Hisano:2015rsa}, the following contributions are included
in the evaluation of the Wilson coefficients at $\mu_{UV}$ with $N_f=5$.\\
\begin{center}
\begin{tabular}{|cc||cc|cc|}
\hline
\multicolumn{2}{|c||}{}&\multicolumn{2}{|c|}{Higgs}&\multicolumn{2}{|c|}{Box}\\
\hline
Paton&Orators &LO &NLO&LO &NLO\\
\hline
quark & scalar & 1 loop & 2 loop&-&2 loop\\
(1st and 2nd gen.)& twist-2 & -& - &1 loop &2 loop\\
\hline
quark & scalar & 1 loop & 2 loop& 1 loop & 2 loop (neglected)\\
(bottom) & twist-2 & -& - &1 loop &2 loop (neglected)\\
\hline
gluon & scalar& 2 loop & 3 loop& 2 loop & 3 loop\\
(1st and 2nd gen.) &twist-2 & -& - & - &2 loop\\
\hline
gluon &scalar & 2 loop & 3 loop& 2 loop & 3 loop (neglected)\\
(3nd gen.)&twist-2 & -& - & - &2 loop (neglected)\\
\hline
\end{tabular}
\end{center}
~\\
Here, ``$-$'' means that the contributions vanish. The column with ``Higgs''
is for the Higgs exchange diagrams and that with ``box'' is for box diagrams.

\begin{figure}
\begin{center}
\includegraphics[width=8cm]{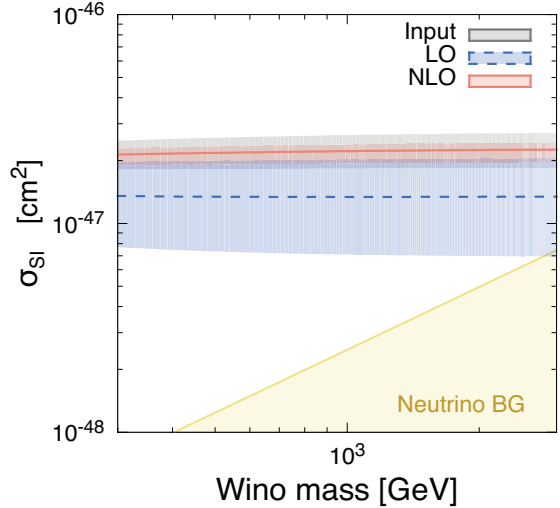}
\end{center}
\caption{SI cross section of SU(2)$_L$ triplet fermions with nucleons.
The (blue) dashed and (red) solid lines are for the leading-order and next-leading order
results, respectively, with corresponding bands for perturbative uncertainties. The gray band is for uncertainty from input errors. In the yellow-shaded area the neutrino
floor dominates the DM signals.
Figure comes from Ref.~\cite{Hisano:2015rsa}.
\label{sicross}}
\end{figure}

It is beyond the scope of this book to show the details of
the calculation. The SI cross section in the limit of $M_X\gg m_W$ is found to be  \cite{Hisano:2015rsa}
\begin{eqnarray}
S^p_{rm SI}&=& 2.3^{+0.2+0.5}_{-0.3-0.4}\times 10^{-47}{\rm cm^2}
\label{winonlo}
\end{eqnarray}
where the first error comes from perturbative uncertainties, mainly
from the choice of $\mu_{UV}$. The second one is from the input uncertainty,
such as the PDFs. It is checked that neglecting the next-leading order
contribution from the third generation quarks does not lead to
significant errors. The SI cross section with protons is shown as a
function of $M_X$ in Fig.~\ref{sicross}. The blue dashed and red
solid lines represent the leading-order and next-leading order results,
respectively, with corresponding bands for perturbative
uncertainties. The gray band is for the uncertainty resulting from the
input error. The yellow shaded area corresponds to the neutrino floor, where
the neutrino background dominates the DM signal. Fortunately, the SI cross
section in Eq.~\ref{winonlo} is larger than the neutrino floor even if
the WIMP mass is around the 3 TeV predicted from thermal production.

The above calculation is applicable to other SU(2)$_L$ multiplets
\cite{Hisano:2015rsa}. If the WIMPs come from SU(2)$_L$ doublet
fermions (Higgsinos in the SUSY SM), the SI cross section with protons
is around $10^{-49}$cm$^2$, which is below the neutrino floor. On the
other hand, larger SU(2)$_L$ multiplets predict larger SI cross
sections. For example, SU(2)$_L$ quintuplet fermions with zero hypercharge
predict $\sigma_{\rm SI}^p\simeq 2\times 10^{-46}$cm$^2$.

\section{Summary}
\label{sec:7}

The DM direct detection experiments are important for tests of the WIMP
DM, and they are a window to new physics at the TeV scale. The
first-generation experiments, LUX, XENON1T, and PandaX-II, are giving the
strict bounds on the SI cross sections of WIMPs with nucleons. In a
few years, the second-generation experiments will begin, LZ,
XENONnT, and PandaX-nT with fiducial volumes $O(1)$~ton. The third-generation experiments with fiducial volumes $O(10)$~ton are also being
planned in order to reach the neutrino floor. Thus, the theoretical
predictions for the detection rates have to be more accurate and
reliable in order to clarify their coverage to survey for WIMP models
and their parameters.

In this lecture, the DM direct detection rates are evaluated with the
effective theories. The effective theory approach is quite useful to
evaluate them in a systematic way. We noted the QCD aspects in the evaluation
of the effective coupling constants of WIMPs with nucleons. It is found 
that they are well controllable, and we may evaluate them at the 
next-leading order of $\alpha_s$ in the effective theory approach.


\appendix
\section{Fock--Schwinger Gauge}
\label{FSgauge}

The Fock--Schwinger gauge is convenient when evaluating the Wilson
coefficients for the effective operators with SU(3)$_C$ field
strengths. In the Feynman gauge, for example, it is tedious to
maintain the tensor structure of the field strength. On the other
hand, when using the Fock--Schwinger gauge, we may introduce the
propagators of colored particles in the background of the gluon field
strength, or the derivatives. Thus, we may maintain the tensor
structure automatically. This review is based on
Ref.~\cite{Novikov:1983gd}.

The Fock--Schwinger gauge is given by
\begin{eqnarray}
 x^\mu A^A_\mu(x)&=&0
\end{eqnarray}
While this gauge fixing is not invariant under translation, the gauge field can be
expanded at $x\simeq 0$ as
\begin{eqnarray}
 A^A_\mu(x)&=&
 \frac1{2\cdot 0!}x^\rho G_{\rho\mu}^A(0)
 + \frac1{3\cdot 1!} x^\alpha x^\rho (D_\alpha G_{\rho\mu}(0))^A
 \nonumber\\
 &&+ \frac1{4\cdot 2!} x^\alpha x^\beta x^\rho (D_\alpha D_\beta G_{\rho\mu}(0))^A
 +\cdots.
 \label{gexpansion}
\end{eqnarray}
The right-hand side in Eq.~\ref{gexpansion} is given with the gauge
covariant terms, such as the gluon field strength or its derivatives.

For the proof, we first need the following relation in the
Fock--Schwinger gauge,  
\begin{eqnarray}
 A_\mu^A(x)&=&\int^1_0 d\alpha~ G_{\rho\mu}^A(\alpha x)~\alpha x^\rho.
\label{ap:fs1}
\end{eqnarray}
It is derived as follows. Using the gauge fixing condition, it is found that 
\begin{eqnarray}
 A_\mu^A(y) &=& -y^\rho\frac{\partial A^A_\rho(y)}{\partial y^\mu}=
 y^\rho G^A_{\rho\mu}(y)-y^\rho \frac{\partial A^A_{\mu}(y)}{\partial y^\rho}.
\end{eqnarray}
By moving the second term in the right-hand side to the left-hand side, and 
 taking $y=\alpha x$, we get 
\begin{eqnarray}
 \alpha x^\rho G^A_{\rho\mu}(\alpha x)&=&
 A_\mu^A(\alpha x)
 +x^\rho \frac{\partial A^A_{\mu}}{\partial x^\rho}\nonumber\\
 &=& \frac{d}{d\alpha} (\alpha A_\mu^A(\alpha x)).
\label{FSgauge1}
\end{eqnarray}
Thus, Eq.~\ref{ap:fs1} is derived.

Next, by expanding the gauge fixing condition at $x\simeq 0$ as
\begin{eqnarray}
 x^\mu(
 A_\mu^A(0)+
 x_{\alpha_1} \partial^{\alpha_1}A_\mu^A(0)+
 \frac12 x_{\alpha_1}x_{\alpha_2} \partial^{\alpha_1}\partial^{\alpha_2}A_\mu^A(0)+\cdots)&=&0,
\end{eqnarray}
we get
\begin{eqnarray}
 x^\mu A_\mu^A(0)&=&0,\nonumber\\
 x^\mu x_{\alpha_1} \partial^{\alpha_1}A_\mu^A(0)&=&0,\nonumber\\
 x^\mu x_{\alpha_1}x_{\alpha_2} \partial^{\alpha_1}\partial^{\alpha_2}A_\mu^A(0)&=&0,\nonumber\\
 \cdots&=&0.
\label{fsgaugeexpand}
\end{eqnarray}
Using these results, it is found that 
\begin{eqnarray}
 x_{\alpha_1}\cdots x_{\alpha_n} \partial^{\alpha_1}\cdots\partial^{\alpha_n}
 G_{\rho\mu}^A(0)
 &=&
 x_{\alpha_1}\cdots x_{\alpha_n} (D^{\alpha_1}\cdots D^{\alpha_n}
 G_{\rho\mu}(0))^A.
 \label{FSgauge2}
\end{eqnarray}
By expanding Eq.~\ref{ap:fs1} around $x\simeq 0$ and applying Eq.~\ref{FSgauge2} to it, we get Eq.~\ref{gexpansion}. In addition, using Eq.~\ref{fsgaugeexpand}, the following equation for fermions (and also for scalars) is also derived,
\begin{eqnarray}
 \psi(x)&=&
\psi(0)
+ x^\alpha D_\alpha\psi(0)
 + \frac12 x^\alpha x^\beta D_\alpha D_\beta\psi(0) 
  +\cdots.
\end{eqnarray}

Next, we derive the quark propagator in the gluon field background, defined as
\begin{eqnarray}
 (i \partial_{\mu}\gamma^{\mu}-g_s\sla{A}-m_q)
 i S(x,y)&=&i\delta^{(4)}(x-y),
 \end{eqnarray}
where $ \sla{A}= A^A_\mu T_A\gamma_\mu$.
The propagator is derived in a perturbative way as
\begin{eqnarray}
 iS(x,y)&=&iS^{(0)}(x-y)
 +\int d^4z i S^{(0)}(x-z)(-i g_s \sla{A}(z)) i S^{(0)}(z-y)
 \nonumber\\
 &&
 +\int d^4z d^4z' i S^{(0)}(x-z)(-i g_s \sla{A}(z)) i S^{(0)}(z-z')
 (-i g_s \sla{A}(z')) i S^{(0)}(z'-y)
 \nonumber\\
 &&+\cdots,
\end{eqnarray}
where $i S^{(0)}(x-y)$ is the propagator under no gluon background.
The gluon background field $A^A_\mu$ may be replaced with the gluon
field strength or its covariant derivatives using Eq.~\ref{gexpansion} in the
Fock--Schwinger gauge. In momentum space, $A^A_\mu$ is given by
\begin{eqnarray}
 A^A_\mu(k)&=&\int A^A_\mu(x) {\rm e}^{i kx} d^4x\nonumber\\
 &=&\frac{i}2 (2\pi)^4 G^A_{\mu\rho}(0)\frac{\partial}{\partial k_\rho}\delta^{(4)}(k)\cdots,
\end{eqnarray}
where the covariant derivatives of $G^A_{\mu\rho}(0)$ are omitted here
since they are irrelevant to the calculation of the DM detection rates in the text.

Using the above results, we get
\begin{eqnarray}
 i S(p) &=&  \int d^4 x ~e^{i p x} \langle
T\{\psi(x)\bar{\psi}(0)\}\rangle \nonumber\\ &=& i S^{(0)}(p)
\nonumber\\ &+& \int d^4 k_1 ~ i S^{(0)}(p)~g_s\gamma^\alpha
\left(\frac12 G_{\alpha\mu} \frac{\partial}{\partial k_{1\mu}}
\delta^{(4)}(k_1)\right) i S^{(0)}(p-k_1) \nonumber\\ &+& \int d^4
k_1d^4 k_2 ~ i S^{(0)}(p) ~g_s\gamma^\alpha \left(\frac12
G_{\alpha\mu} \frac{\partial}{\partial k_{1\mu}}
\delta^{(4)}(k_1)\right) i S^{(0)}(p-k_1) ~g_s\gamma^\beta
\nonumber\\ &&\times \left(\frac12 G_{\beta\nu}
\frac{\partial}{\partial k_{2\nu}} \delta^{(4)}(k_2)\right) i
S^{(0)}(p-k_1-k_2) + \cdots ~,
\end{eqnarray}
where $i S^{(0)}(p)=i/(\sla{p}-m_q)$ and $G_{\mu\nu}\equiv G_{\mu\nu}^AT_A$. 

Since translation invariance is broken due to the gauge fixing,
$S(0,x)$ does not have the same form as $S(x,0)$,
\begin{eqnarray}
i \tilde{S}(p) &\equiv& \int d^4 x~
e^{-i p x} \langle T\{\psi(0)\bar{\psi}(x)\}\rangle \nonumber\\ &=& i
S^{(0)}(p) \nonumber\\ &+& \int d^4 k_1 ~ i
S^{(0)}(p+k_1)~g_s\gamma^\alpha \left(\frac12 G_{\alpha\mu}
\frac{\partial}{\partial k_{1\mu}} \delta^{(4)}(p)\right) i S^{(0)}(p)
\nonumber\\ &+& \int d^4 k_1d^4 k_2 ~ i
S^{(0)}(p+k_1+k_2)~g_s\gamma^\alpha \left(\frac12 G_{\alpha\mu}
\frac{\partial}{\partial k_{2\mu}} \delta^{(4)}(k_2)\right)
\nonumber\\ &&\times i S^{(0)}(p+k_1)~g_s \gamma^\beta \left(\frac12 g
G_{\beta\nu} \frac{\partial}{\partial k_{1\nu}}
\delta^{(4)}(k_1)\right) i S^{(0)}(p) + \cdots ~.
\end{eqnarray}

The colored scalar propagator in the gluon background is also derived 
as 
\begin{eqnarray}
i \Delta(p) &\equiv& \int 
d^4 x ~e^{i p x} 
\langle T\{\phi(x)\phi^\dagger(0)\}\rangle
\nonumber\\
&=& i \Delta^{(0)}(p) \nonumber\\
&+& \int d^4 k_1 ~ i \Delta^{(0)}(p)
~g_s(2p-k_1)^\alpha
\left(\frac12 G_{\alpha\mu} 
\frac{\partial}{\partial k_{1\mu}} 
\delta^{(4)}(k_1)\right)
i \Delta^{(0)}(p-k_1) \nonumber\\
&+& \int d^4 k_1d^4 k_2 ~ i \Delta^{(0)}(p)
~g_s(2p-k_1)^\alpha
\left(\frac12 G_{\alpha\mu} 
\frac{\partial}{\partial k_{1\mu}} 
\delta^{(4)}(k_1)\right)
i \Delta^{(0)}(p-k_1) 
\nonumber\\
&&\times 
g_s(2p-2k_1-k_2)^\beta
\left(\frac12 G_{\beta\nu} 
\frac{\partial}{\partial k_{2\nu}} 
\delta^{(4)}(k_2)\right)
i \Delta^{(0)}(p-k_1-k_2)
\nonumber \\
&+& \int d^4 k_1d^4 k_2 ~ i \Delta^{(0)}(p)
(-ig_s^2)
\left(\frac12 G_{\alpha\mu} 
\frac{\partial}{\partial k_{1\mu}} 
\delta^{(4)}(k_1)\right)
\left(\frac12 G^\alpha_{~\nu} 
\frac{\partial}{\partial k_{2\nu}} 
\delta^{(4)}(k_2)\right)
\nonumber \\
&&\times 
i \Delta^{(0)}(p-k_1-k_2)\ ,
\end{eqnarray}
\begin{eqnarray}
i \tilde{\Delta}(p) &\equiv& \int 
d^4 x ~e^{-i p x} 
\langle T\{\phi(0)\phi^\dagger(x)\}\rangle
\nonumber\\
&=& i \Delta^{(0)}(p) \nonumber\\
&+& \int d^4 k_1 ~ i \Delta^{(0)}(p+k_1)
~g_s(2p+k_1)^\alpha
\left(\frac12 G_{\alpha\mu} 
\frac{\partial}{\partial k_{1\mu}} 
\delta^{(4)}(k_1)\right)
i \Delta^{(0)}(p) \nonumber\\
&+& \int d^4 k_1d^4 k_2 ~ i \Delta^{(0)}(p+k_1+k_2)
~g_s(2p+2k_1+k_2)^\alpha
\left(\frac12 G_{\alpha\mu} 
\frac{\partial}{\partial k_{1\mu}} 
\delta^{(4)}(k_2)\right)
\nonumber\\
&&\times
i \Delta^{(0)}(p+k_1) ~
g_s (2p+k_1)^\beta 
\left(\frac12 G_{\beta\nu} 
\frac{\partial}{\partial k_{2\nu}} 
\delta^{(4)}(k_1)\right)
i \Delta^{(0)}(p)
\nonumber \\
&+& \int d^4 k_1d^4 k_2 ~ i \Delta^{(0)}(p+k_1+k_2)
\nonumber \\
&&\times 
(-ig_s^2)
\left(\frac12 G_{\alpha\mu} 
\frac{\partial}{\partial k_{1\mu}} 
\delta^{(4)}(k_1)\right)
\left(\frac12 G^\alpha_{~\nu} 
\frac{\partial}{\partial k_{2\nu}} 
\delta^{(4)}(k_2)\right)
i \Delta^{(0)}(p)\ ,
\end{eqnarray}
where $i \Delta^{(0)}(p)=i/(p^2-m^2)$. The scalar propagators are 
reduced to a more convenient form for practical usage as \cite{Hisano:2010ct}
\begin{eqnarray}
i \Delta(p) 
&=& i \Delta^{(0)}(p) \nonumber\\
\nonumber \\
&+& \int d^4 k_1d^4 k_2 ~ i \Delta^{(0)}(p)
(-ig_s^2)
\left(\frac12 G_{\alpha\mu} 
\frac{\partial}{\partial k_{1\mu}} 
\delta^{(4)}(k_1)\right)
\left(\frac12 G^\alpha_{~\nu} 
\frac{\partial}{\partial k_{2\nu}} 
\delta^{(4)}(k_2)\right)
\nonumber \\
&&\times 
i \Delta^{(0)}(p-k_1-k_2) 
\nonumber\\
&&+\cdots,
\nonumber\\
i \tilde{\Delta}(p) 
&=& i \Delta^{(0)}(p) \nonumber\\
&+& \int d^4 k_1d^4 k_2 ~ i \Delta^{(0)}(p+k_1+k_2)
\nonumber \\
&&\times 
(-ig_s^2)
\left(\frac12 G_{\alpha\mu} 
\frac{\partial}{\partial k_{1\mu}} 
\delta^{(4)}(k_1)\right)
\left(\frac12 G^\alpha_{~\nu} 
\frac{\partial}{\partial k_{2\nu}} 
\delta^{(4)}(k_2)\right)
i \Delta^{(0)}(p) \nonumber\\
&&\cdots.
\end{eqnarray}

The scalar
operator of the gluon $G_{\mu\nu}^AG^{A\mu\nu}$ is projected out from the
bilinear term of the gluon field strength as
\begin{eqnarray}
G_{\alpha\mu}^AG_{\beta\nu}^A
&=&
\frac{1}{12}
G_{\rho\sigma}^AG^{A\rho\sigma}
(g_{\alpha\beta}g_{\mu\nu}
-g_{\alpha\nu}g_{\beta\mu}) 
\nonumber\\
&-&\frac12 g_{\alpha\beta} {O}_{\mu\nu}^g
-\frac12 g_{\mu\nu} {O}_{\alpha\beta}^g
-\frac12 g_{\alpha\nu} {O}_{\beta\mu}^g
-\frac12 g_{\beta\mu} {O}_{\alpha\nu}^g
\nonumber\\
&+& {O}_{\alpha\mu\beta\nu}^g \ ,
\end{eqnarray}
where ${O}_{\mu\nu}^g$ is the twist-2 operator of gluon
(Eq.~(\ref{twist2op})) and ${O}_{\alpha\mu\beta\nu}^g$ is given as 
 \begin{eqnarray}
 {O}_{\alpha\mu\beta\nu}^g &\equiv&
  G_{\alpha\mu}^A G_{\beta\nu}^A
  \nonumber\\
  &-&\frac12 g_{\alpha\beta} G_{~~\mu}^{A\rho}G_{\rho\nu}^{A}
  -\frac12 g_{\mu\nu} G_{~~\alpha}^{A\rho}G_{\rho\beta}^{A} +\frac12
  g_{\alpha\nu} G_{~~\beta}^{A\rho}G_{\rho\mu}^{A} +\frac12
  g_{\beta\mu} G_{~~\alpha}^{A\rho}G_{\rho\nu}^{A}
  \nonumber\\
  &+&\frac16 G_{\rho\sigma}^AG^{A\rho\sigma}
  (g_{\alpha\beta}g_{\mu\nu} -g_{\alpha\nu}g_{\beta\mu}) \ .
\end{eqnarray}

\begin{figure}
  \begin{center}
   \begin{picture}(150,50)(0,0)
    \ArrowLine(20,25)(45,45)
    \ArrowLine(45,45)(45,5)
    \ArrowLine(45,5)(20,25)
    \Gluon(45,45)(60,45){2}{2}
    \Gluon(45,5)(60,5){2}{2}
    \ArrowLine(100,25)(125,45)
    \ArrowLine(125,45)(125,5)
    \ArrowLine(125,5)(100,25)
    \Gluon(125,45)(140,5){2}{4}
    \Gluon(125,5)(140,45){2}{4}
   \end{picture}
\end{center}
\caption{Integrating out the heavy quark. }
\label{heavyquarkintegral1}
 \end{figure}
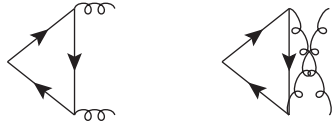

 Finally, we show Eq.~\ref{heavyquarkintegral} by integrating out heavy
 quarks in the Fock--Schwinger gauge. In the text, we derived this using the trace anomaly in QCD. Here,
 we will show it in a diagrammatic way. As in Fig.~\ref{heavyquarkintegral1}, two
 diagrams contribute there. However, in the Fock--Schwinger gauge, it is
 given by the trace of the quark propagator as 
 \begin{eqnarray}
  i M= -im_Q\int\frac{d^4p}{(4\pi)^4}{\rm Tr}_{C+L}[iS(p)]=i\frac{\alpha_s}{12\pi}
  G_{\rho\sigma}^AG^{A\rho\sigma}.
\end{eqnarray}
While the Fock--Schwinger gauge is not invariant under translation, the
invariance is recovered in the gauge-invariant results. The above
calculation is quite easy compared with the other gauges, such as the
Feynman gauge.

\thebibliography{0}

\bibitem{Bertone:2004pz}
For reviews of dark matter, 
 G.~Jungman, M.~Kamionkowski and K.~Griest,
 Phys.\ Rept.\ {\bf 267} (1996) 195
 [hep-ph/9506380];\\
K.~A.~Olive,
 ``TASI lectures on dark matter,''
 astro-ph/0301505;\\
 G.~Bertone, D.~Hooper and J.~Silk,
 Phys.\ Rept.\ {\bf 405} (2005) 279
 [hep-ph/0404175];\\
J.~L.~Feng,
 Ann.\ Rev.\ Astron.\ Astrophys.\ {\bf 48} (2010) 495
 [arXiv:1003.0904 [astro-ph.CO]].

\bibitem{Undagoitia:2015gya}
 T.~Marrodán Undagoitia and L.~Rauch,
 J.\ Phys.\ G {\bf 43} (2016) no.1, 013001 
 [arXiv:150\\9.08767 [physics.ins-det]].

\bibitem{Hisano:2015bma}
 J.~Hisano, R.~Nagai and N.~Nagata,
 JHEP {\bf 1505} (2015) 037
 [arXiv:1502.02244 [hep-ph]].

\bibitem{Hisano:2015rsa}
 J.~Hisano, K.~Ishiwata and N.~Nagata,
 JHEP {\bf 1506} (2015) 097
 [arXiv:1504.00915 [hep-ph]].

\bibitem{Ade:2013zuv} 
 P.~A.~R.~Ade {\it et al.} [Planck Collaboration],
 Astron.\ Astrophys.\ (2014)
 [arXiv:1303.5076 [astro-ph.CO]].

\bibitem{SUSYDM}
For reviews of SUSY SM and SUSY DM, 
 G.~Jungman, M.~Kamionkowski and K.~Griest in \cite{Bertone:2004pz};\\
S.~P.~Martin,
 Adv.\ Ser.\ Direct.\ High Energy Phys.\ {\bf 21} (2010) 1
  [Adv.\ Ser.\ Direct.\ High Energy Phys.\ {\bf 18} (1998) 1]
 [hep-ph/9709356].

\bibitem{Appelquist:2000nn}
 T.~Appelquist, H.~C.~Cheng and B.~A.~Dobrescu,
 Phys.\ Rev.\ D {\bf 64} (2001) 035002
 [hep-ph/0012100].

\bibitem{Cheng:2002ej}
 H.~C.~Cheng, J.~L.~Feng and K.~T.~Matchev,
 Phys.\ Rev.\ Lett.\ {\bf 89} (2002) 211301
 [hep-ph/0207125];\\
H.~C.~Cheng, K.~T.~Matchev and M.~Schmaltz,
 Phys.\ Rev.\ D {\bf 66} (2002) 036005
 [hep-ph/0204342].

\bibitem{boltmaneq}
The Boltzmann equation for WIMPs is given in standard textbooks, such as 
E.~W.~Kolb and M.~S.~Turner,
``The Early Universe.''

\bibitem{Edsjo:1997bg}
 J.~Edsjo and P.~Gondolo,
 Phys.\ Rev.\ D {\bf 56} (1997) 1879
 [hep-ph/9704361].

\bibitem{Hisano:2006nn}
 J.~Hisano, S.~Matsumoto, M.~Nagai, O.~Saito and M.~Senami,
 Phys.\ Lett.\ B {\bf 646} (2007) 34
 [hep-ph/0610249];\\
M.~Cirelli, A.~Strumia and M.~Tamburini,
 Nucl.\ Phys.\ B {\bf 787} (2007) 152 
 [arXiv:0706.4071 [hep-ph]].

\bibitem{TheFermi-LAT:2015kwa}
 M.~Ajello {\it et al.} [Fermi-LAT Collaboration],
 Astrophys.\ J.\ {\bf 819} (2016) no.1, 44
 [arXiv:1511.02938 [astro-ph.HE]].

\bibitem{Ahnen:2016qkx}
 M.~Ackermann {\it et al.} [Fermi-LAT Collaboration],
 Phys.\ Rev.\ Lett.\ {\bf 115} (2015) no.23, 231301
 [arXiv:1503.02641 [astro-ph.HE]];\\
 M.~L.~Ahnen {\it et al.} [MAGIC and Fermi-LAT Collaborations],
 JCAP {\bf 1602} (2016) no.02, 039
 [arXiv:1601.06590 [astro-ph.HE]].

\bibitem{Abramowski:2011hc}
 A.~Abramowski {\it et al.} [H.E.S.S. Collaboration],
 Phys.\ Rev.\ Lett.\ {\bf 106} (2011) 161301
 [arXiv:1103.3266 [astro-ph.HE]];\\
 M.~L.~Ahnen {\it et al.} in \cite{Ahnen:2016qkx}.

\bibitem{Aguilar:2016kjl}
 M.~Aguilar {\it et al.} [AMS Collaboration],
 Phys.\ Rev.\ Lett.\ {\bf 117} (2016) no.9, 091103.

\bibitem{Aartsen:2012kia}
 M.~G.~Aartsen {\it et al.} [IceCube Collaboration],
 Phys.\ Rev.\ Lett.\ {\bf 110} (2013) no.13, 131302
 [arXiv:1212.4097 [astro-ph.HE]].

\bibitem{Choi:2015ara}
 K.~Choi {\it et al.} [Super-Kamiokande Collaboration],
 Phys.\ Rev.\ Lett.\ {\bf 114} (2015) no.14, 141301
 [arXiv:1503.04858 [hep-ex]].

\bibitem{Hisano:2003ec}
 J.~Hisano, S.~Matsumoto and M.~M.~Nojiri,
 Phys.\ Rev.\ Lett.\ {\bf 92} (2004) 031303
 [hep-ph/0307216];\\
 J.~Hisano, S.~Matsumoto, M.~M.~Nojiri and O.~Saito,
  Phys.\ Rev.\ D {\bf 71} (2005) 063528
  [hep-ph/0412403];\\
J.~Hisano, S.~Matsumoto, O.~Saito and M.~Senami,
 Phys.\ Rev.\ D {\bf 73} (2006) 055004
 [hep-ph/0511118].

\bibitem{Aprile:2017iyp}
 E.~Aprile {\it et al.} [XENON Collaboration],
 Phys.\ Rev.\ Lett.\ {\bf 119} (2017) no.18, 181301
 [arXiv:1705.06655 [astro-ph.CO]].

\bibitem{Cui:2017nnn}
 X.~Cui {\it et al.} [PandaX-II Collaboration],
 Phys.\ Rev.\ Lett.\ {\bf 119} (2017) no.18, 181302
 [arXiv:1708.06917 [astro-ph.CO]].

\bibitem{Fan:2010gt}
 J.~Fan, M.~Reece and L.~T.~Wang,
 JCAP {\bf 1011} (2010) 042
 [arXiv:1008.1591 [hep-ph]].

\bibitem{Drees:1993bu}
 M.~Drees and M.~Nojiri,
 Phys.\ Rev.\ D {\bf 48} (1993) 3483
 [hep-ph/9307208].

\bibitem{Lewin:1995rx}
 J.~D.~Lewin and P.~F.~Smith,
 Astropart.\ Phys.\ {\bf 6} (1996) 87.

\bibitem{Read:2014qva}
 J.~I.~Read,
 J.\ Phys.\ G {\bf 41} (2014) 063101
 [arXiv:1404.1938 [astro-ph.GA]].

\bibitem{Akerib:2016vxi}
 D.~S.~Akerib {\it et al.} [LUX Collaboration],
 Phys.\ Rev.\ Lett.\ {\bf 118} (2017) no.2, 021303
 [arXiv:1608.07648 [astro-ph.CO]].

\bibitem{Gutlein:2010tq}
 A.~Gutlein {\it et al.},
 Astropart.\ Phys.\ {\bf 34} (2010) 90
 [arXiv:1003.5530 [hep-ph]];\\
 J.~Billard, L.~Strigari and E.~Figueroa-Feliciano,
 Phys.\ Rev.\ D {\bf 89} (2014) no.2, 023524
 [arXiv:1307.5458 [hep-ph]];\\
F.~Ruppin, J.~Billard, E.~Figueroa-Feliciano and L.~Strigari,
 Phys.\ Rev.\ D {\bf 90} (2014) no.8, 083510
 [arXiv:1408.3581 [hep-ph]].

\bibitem{Hill:2013hoa}
 R.~J.~Hill and M.~P.~Solon,
 Phys.\ Rev.\ Lett.\ {\bf 112} (2014) 211602
 [arXiv:1309.4092 [hep-ph]].

\bibitem{Abdel-Rehim:2016won}
 A.~Abdel-Rehim {\it et al.} [ETM Collaboration],
 Phys.\ Rev.\ Lett.\ {\bf 116} (2016) no.25, 252001
 [arXiv:1601.01624 [hep-lat]].

\bibitem{Shifman:1978zn}
 M.~A.~Shifman, A.~I.~Vainshtein and V.~I.~Zakharov,
 Phys.\ Lett.\ {\bf 78B} (1978) 443.

\bibitem{Cho:1994yu}
L.~Vecchi, [arXiv:1312.5695];\\
 P.~L.~Cho and E.~H.~Simmons,
 Phys.\ Rev.\ D {\bf 51} (1995) 2360
 [hep-ph/9408206].

\bibitem{Peskin:1995ev}
 M.~E.~Peskin and D.~V.~Schroeder,
 ``An Introduction to quantum field theory.''

\bibitem{Schwartz:2013pla}
 M.~D.~Schwartz,
 ``Quantum Field Theory and the Standard Model.''

\bibitem{Collins:1981uw}
 J.~C.~Collins and D.~E.~Soper,
 Nucl.\ Phys.\ B {\bf 194} (1982) 445.

\bibitem{Soper:1996sn}
 D.~E.~Soper,
 Nucl.\ Phys.\ Proc.\ Suppl.\ {\bf 53} (1997) 69
 [hep-lat/9609018].

\bibitem{Owens:2012bv}
 J.~F.~Owens, A.~Accardi and W.~Melnitchouk,
 Phys.\ Rev.\ D {\bf 87} (2013) no.9, 094012
 [arXiv:1212.1702 [hep-ph]].

\bibitem{Adams:1995ufa}
 D.~Adams {\it et al.} [Spin Muon Collaboration],
 Phys.\ Lett.\ B {\bf 357} (1995) 248.
 
\bibitem{Floratos:1978ny}
 E.~G.~Floratos, D.~A.~Ross and C.~T.~Sachrajda,
 Nucl.\ Phys.\ B {\bf 152} (1979) 493;\\
A.~Gonzalez-Arroyo and C.~Lopez,
 Nucl.\ Phys.\ B {\bf 166} (1980) 429.

\bibitem{Griest:1989wd}
 K.~Griest and M.~Kamionkowski,
 Phys.\ Rev.\ Lett.\ {\bf 64} (1990) 615.

\bibitem{Ibe:2012sx}
 M.~Ibe, S.~Matsumoto and R.~Sato,
 Phys.\ Lett.\ B {\bf 721} (2013) 252
 [arXiv:1212.5989 [hep-ph]].

\bibitem{Giudice:1998xp}
L.~Randall and R.~Sundrum,
 Nucl.\ Phys.\ B {\bf 557} (1999) 79
 [hep-th/9810155];\\
 G.~F.~Giudice, M.~A.~Luty, H.~Murayama and R.~Rattazzi,
 JHEP {\bf 9812} (1998) 027
 [hep-ph/9810442].

\bibitem{Hisano:2012wm}
 J.~Hisano, K.~Ishiwata and N.~Nagata,
 Phys.\ Rev.\ D {\bf 87} (2013) 035020
 [arXiv:1210.5985 [hep-ph]].

\bibitem{Hisano:2010fy}
 J.~Hisano, K.~Ishiwata and N.~Nagata,
 Phys.\ Lett.\ B {\bf 690} (2010) 311
 [arXiv:1004.4090 [hep-ph]].

\bibitem{Hisano:2004pv}
 J.~Hisano, S.~Matsumoto, M.~M.~Nojiri and O.~Saito,
 Phys.\ Rev.\ D {\bf 71} (2005) 015007
 [hep-ph/0407168].

\bibitem{Hisano:2011cs}
 J.~Hisano, K.~Ishiwata, N.~Nagata and T.~Takesako,
 JHEP {\bf 1107} (2011) 005
 [arXiv:1104.0228 [hep-ph]].

\bibitem{Hisano:2014kua}
 J.~Hisano, D.~Kobayashi, N.~Mori and E.~Senaha,
 Phys.\ Lett.\ B {\bf 742} (2015) 80
 [arXiv:1410.3569 [hep-ph]].


 \bibitem{Broadhurst:1994qj}
 D.~J.~Broadhurst, P.~A.~Baikov, V.~A.~Ilyin, J.~Fleischer, O.~V.~Tarasov and V.~A.~Smirnov,
 Phys.\ Lett.\ B {\bf 329} (1994) 103
 [hep-ph/9403274].

\bibitem{Bardeen:1978yd}
 W.~A.~Bardeen, A.~J.~Buras, D.~W.~Duke and T.~Muta,
 Phys.\ Rev.\ D {\bf 18} (1978) 3998.

\bibitem{Djouadi:2000ck}
 A.~Djouadi and M.~Drees,
 Phys.\ Lett.\ B {\bf 484} (2000) 183
 [hep-ph/0004205].

\bibitem{Novikov:1983gd}
 V.~A.~Novikov, M.~A.~Shifman, A.~I.~Vainshtein and V.~I.~Zakharov,
 Fortsch.\ Phys.\ {\bf 32} (1984) 585.
 
\bibitem{Hisano:2010ct}
 J.~Hisano, K.~Ishiwata and N.~Nagata,
 Phys.\ Rev.\ D {\bf 82} (2010) 115007
 [arXiv:1007.2601 [hep-ph]].

\endthebibliography
\end{document}